\documentclass[acmsmall,screen]{acmart}
\setcopyright{none}
\renewcommand\footnotetextcopyrightpermission[1]{} % removes footnote with conference information in first column
\usepackage{makecell}
\usepackage{multirow}
\usepackage{xspace}
\usepackage{amsmath}
\usepackage[most]{tcolorbox}
\usepackage{soul}
\usepackage{tikz}
\usepackage{circledsteps}
\usepackage{subcaption}

\newcommand{\bighead}[1]{\par\smallskip\noindent\textbf{#1.}}
\newcommand{\head}[1]{\par\smallskip\noindent\textit{\underline{#1}}}
\newcommand{\rev}[1]{\textcolor{black}{#1}}

\newcommand\bluecircled[1]{%
  \Circled{#1}}

\newcommand{\skybluecircled}[1]{
  \Circled{#1}}

\newcommand{\redcircled}[1]{
  \Circled{#1}}

\newcommand{\greencircled}[1]{
  \Circled{#1}}

\newcommand{\tealcircled}[1]{
  \Circled{#1}}

\newcommand{\mauvecircled}[1]{
  \Circled{#1}}

\newboolean{showcomments}
\setboolean{showcomments}{true}

\ifthenelse{\boolean{showcomments}}
{\newcommand{\nbc}[3]{
 {\colorbox{#3}{\bfseries\sffamily\scriptsize\textcolor{white}{#1}}}
 {\textcolor{#3}{\sf\small$\blacktriangleright$\textit{#2}$\blacktriangleleft$}}
 }
}
{\newcommand{\nbc}[3]{}
 }

\newcommand{\fm}{FM\xspace}
\newcommand{\fms}{FMs\xspace}
\newcommand{\toggle}{\textsf{Toggle}\xspace}
\newcommand{\alpharepair}{\textsf{AlphaRepair}\xspace}
\newcommand{\core}{\textsf{CORE}\xspace}
\newcommand{\localizeagent}{\textsf{LOCALIZEAGENT}\xspace}
\newcommand{\staticanalyzer}{\textsf{Timon}\xspace}

\newcommand{\domainspecificlanguage}{domain-specific language\xspace}
\newcommand{\workflow}{FM-generated DSL workflow\xspace}
\newcommand{\dsl}{DSL\xspace}
\newcommand{\json}{JSON\xspace}
\newcommand{\defectcount}{$20$\xspace}
\newcommand{\detectcount}{nine\xspace}

\newcommand{\undefvar}{unreachable variables\xspace}
\newcommand{\unuse}{unused variables\xspace}
\newcommand{\badtype}{incorrect data type propagation\xspace}
\newcommand{\badoutput}{incorrect outputs\xspace}
\newcommand{\badjson}{unparsable JSON\xspace}
\newcommand{\baddsl}{invalid DSL\xspace}
\newcommand{\badexp}{malformed expressions\xspace}
\newcommand{\hallucinatedskills}{hallucinated skills\xspace}
\newcommand{\badparameter}{skills with defective parameters\xspace}
\newcommand{\cundefvar}{Unreachable variables\xspace}
\newcommand{\cunuse}{Unused variables\xspace}
\newcommand{\cbadtype}{Incorrect data type propagation\xspace}
\newcommand{\cbadoutput}{Incorrect outputs\xspace}
\newcommand{\cbadjson}{Unparsable JSON\xspace}
\newcommand{\cbaddsl}{Invalid DSL\xspace}
\newcommand{\cbadexp}{Malformed expressions\xspace}
\newcommand{\challucinatedskills}{Hallucinated skills\xspace}
\newcommand{\cbadparameter}{Skills with defective parameters\xspace}
\newcommand{\pydantic}{Pydantic\xspace}

\newcommand{\openai}{OpenAI\xspace}
\newcommand{\meta}{Meta\xspace}
\newcommand{\an}{Anthropic\xspace}
\newcommand{\google}{Google\xspace}
\newcommand{\llama}{Llama\xspace}
\newcommand{\gpt}{GPT\xspace}
\newcommand{\gemini}{Gemini\xspace}
\newcommand{\qwen}{Qwen\xspace}
\newcommand{\usedef}{use-def chain\xspace}
\newcommand{\defuse}{def-use chain\xspace}
\newcommand{\usedefdefuse}{use-def and def-use chains\xspace}

\newcommand{\taskaction}{task-corresponding action\xspace}

\newcommand{\correctionmodule}{\textsf{Pumbaa}\xspace}
\newcommand{\abst}{abstract syntax tree\xspace}
\newcommand{\totalsize}{$55$\xspace}
\newcommand{\samplesize}{$40$\xspace}

\newcommand{\truepositives}{true positives\xspace}
\newcommand{\falsepositives}{false positives\xspace}
\newcommand{\falsenegatives}{false negatives\xspace}
\newcommand{\tp}{TP\xspace}
\newcommand{\fp}{FP\xspace}
\newcommand{\fn}{FN\xspace} 
\newcommand{\precision}{precision\xspace}
\newcommand{\recall}{recall\xspace}
\newcommand{\maximumcountofrepairattempts}{maximum count of repair attempts\xspace}
\newcommand{\attemptcount}{ten\xspace}
\newcommand{\passk}{\textit{pass@k}\xspace}
\newcommand{\reqone}{\textbf{RQ1:} What are the characteristics of defects in \workflow{}s?\xspace}
\newcommand{\gallaba}{interactive, multi-agent requirement refiner framework\xspace}
\newcommand{\alignmind}{\textsc{AlignMind}\xspace}
\newcommand{\wfdsl}{\textsc{Workflow\dsl}\xspace}

\newcommand{\reqtwo}{\textbf{RQ2:} How accurately can \workflow{} defects be detected using static analysis?\xspace}
\newcommand{\reqthree}{\textbf{RQ3:} How effective are FMs at repairing \workflow{} defects when provided with static analysis feedback?\xspace}
\newcommand{\rqoneone}{RQ1: Observation 1\xspace}
\newcommand{\rqonetwo}{RQ1: Obesrvation 2\xspace}
\newcommand{\justrqtwo}{RQ2\xspace}
\newcommand{\justrqthree}{RQ3\xspace}
\newcommand{\opencoding}{open-coding\xspace}
\newcommand{\opencoded}{open-coded\xspace}

\newcommand{\practicalimplication}{\textit{Implication}}
\newcommand{\csmt}{\textit{Satisfiability Modulo Theories} (SMT) solvers\xspace}
\newcommand{\smt}{SMT}
\newcommand{\dspy}{\textsf{DSPy Optimizer}\xspace}
\newcommand{\anthropic}{\textsf{Anthropic Prompt Improver}\xspace}
\newcommand{\zthree}{\textsf{Z3 Theorem Prover}\xspace}
\newcommand{\openagi}{OpenAGI\xspace}
\copyrightyear{2026}
\acmYear{2026}
\acmJournal{TOSEM}
\acmVolume{}
\acmNumber{}
\acmArticle{}
\acmMonth{03}

\begin{document}

\title{Towards Reliable Generation of Executable Workflows by Foundation Models}
% \title{Automatic Synthesis and Repair of General-Purpose Workflow Programs by Foundation Models}
% \title{From Natural Language to Executable Workflows: Automatic Program Synthesis and Repair by Foundation Models}


\author{Sogol Masoumzadeh}
\orcid{0009-0002-6626-1919}
\email{sogol.masoumzadeh@mail.mcgill.ca}
\affiliation{%
  \institution{McGill University, Montréal}
  \state{QC}
  \country{Canada}}
  
\author{Keheliya Gallaba}
\orcid{0000-0002-5880-5114}
\email{keheliya.gallaba@huawei.com}
\affiliation{%
  \institution{Centre for Software Excellence, Huawei Canada}
  \state{ON}
  \country{Canada}}

\author{Dayi Lin}
\orcid{0000-0002-4034-6650}
\email{dayi.lin@huawei.com}
\affiliation{%
  \institution{Centre for Software Excellence, Huawei Canada}
  \state{ON}
  \country{Canada}}

\author{Ahmed E. Hassan}
\orcid{0000-0001-7749-5513}
\email{ahmed@queensu.ca}
\affiliation{%
  \institution{Queen's University}
  \state{ON}
  \country{Canada}}

\renewcommand{\shortauthors}{Masoumzadeh et al.}

\begin{abstract}

Recent advancements in Foundation Models (\fms) have demonstrated significant progress in processing complex natural language to perform intricate tasks. Successfully executing these tasks often requires orchestrating calls to \fms alongside other software components. However, manually decomposing a task into a coherent sequence of 
smaller, logically aggregated steps, commonly referred to as workflows, demands considerable effort and specialized domain knowledge. While \fms can assist in generating such workflows specified in \domainspecificlanguage{}s (\dsl{}s), achieving accuracy and reliability in this process remains a challenge.
 
We introduce a framework that leverages static analysis feedback to enable FMs to detect and repair defects in the DSL-based workflows they generate.
We begin by presenting \rev{an initial taxonomy} %the first-ever taxonomy 
of incidences of defects in \workflow{}s, categorizing them into \rev{\defectcount} distinct types. 
Furthermore, we observe a high prevalence of defects across \workflow{}s, with \rev{$89.23$\%} %$87.27$\% 
of the studied instances containing at least one defect. This high prevalence underscores the magnitude of the problem and the necessity for mitigation strategies. Following this, we demonstrate that \detectcount types of these defects can be effectively identified through static analysis of the workflows. For this purpose, we develop \staticanalyzer, the first-of-its-kind static analyzer specifically designed for \workflow{}s. Finally, we show that by incorporating feedback from \staticanalyzer, we can guide \correctionmodule, an \fm-based tool, to repair the detected defect incidences. 
By systematically detecting and repairing defects, our work provides a crucial step towards the reliable and automated generation of executable workflows from natural language requirements.
\end{abstract}

\begin{CCSXML}
<ccs2012>
   <concept>
       <concept_id>10010147.10010257</concept_id>
       <concept_desc>Computing methodologies~Machine learning</concept_desc>
       <concept_significance>500</concept_significance>
       </concept>
   <concept>
       <concept_id>10011007.10011074.10011092</concept_id>
       <concept_desc>Software and its engineering~Software development techniques</concept_desc>
       <concept_significance>500</concept_significance>
       </concept>
 </ccs2012>
\end{CCSXML}



%% frame
\newtcolorbox{myframe}[1][]{
  enhanced,
  sharp corners, % Ensures no rounded edges
  colback=white,
  boxrule=0.6pt,
  colback=gray!6, % Very light gray background for subtle shading
  #1
}
%% listing defintion
\lstdefinelanguage{wf}{
    morekeywords={workflow, inputs, tasks, id, if, condition, then, else},
    sensitive=true,
    morecomment=[l]{\#},
    morestring=[b]",
    morestring=[b]',
}
\newtcolorbox{wf}{
    colback=gray!10, % Background color
    colframe=black,  % Frame color
    boxsep=0pt,      % Padding
    arc=0pt,         % Corner radius
    boxrule=0.5pt    % Frame width
}

% Define style for IaC code
\lstset{
    basicstyle=\footnotesize\ttfamily,
    keywordstyle=\color{blue},
    stringstyle=\color{level3},
    breaklines=true,
    numbers=left,        % Add line numbers
    numberstyle=\tiny\color{black},   % Style for line numbers
    numbersep=5pt,       % Adjust spacing between numbers and code
    showstringspaces=false,
    escapeinside={*@}{@*},
            literate=
                  {\{}{{\color{level1}\{}}1
                  {\}}{{\color{level1}\}}}1
                  {[}{{\color{level2}[}}1
                  {]}{{\color{level2}]}}1
                  {,}{{\color{black},}}1
         % literate={\{}{{\textcolor{level1}{\{}}}1
         %     {\}}{{\textcolor{level1}{\}}}}1
         %     {[}{{\textcolor{level2}{[}}}1
         %     {]}{{\textcolor{level2}{]}}}1
         %     {,}{{\textcolor{black}{,}}}1
} 
% colors of the listing
% Define colors for different bracket levels
\definecolor{level1}{RGB}{255, 165, 80}  % Porange (Pinkish-Orange)    % Yellow
\definecolor{level2}{RGB}{255, 0, 255}    % Magenta
\definecolor{level3}{RGB}{0, 150, 0}      % Green

\maketitle

\section{Introduction}
\label{sec:introduction}

Foundation Models (FMs), which are large-scale deep learning models trained on vast corpora, have recently demonstrated remarkable advancements in parsing, interpreting, and generating natural language to address complex tasks. From automated code generation~\cite{wei2024magicoder, luo2023wizardcoder, starcoder2} to question~answering~\cite{ma-etal-2023-chain, zhang2024endtoend}, and large-scale document summarization, FMs exhibit an impressive capacity to handle diverse and intricate requirements. However, fully leveraging this capacity in software development scenarios often demands the orchestration of multiple calls to FMs alongside various software components and services. This orchestration is commonly specified as a \textit{workflow}. In this context, a workflow is a sequence of smaller, atomic steps that are logically aggregated to create an automatable end-to-end process~\cite{russell2016workflow}, wherein each step represents an invocation of an FM or a supporting software element (e.g., databases, external APIs, or transformation scripts). 

% To enable autonomy and flexibility, 
To reduce manual development effort while enabling autonomy and flexibility, developers have considered using conventional enterprise business process languages such as BPEL and BPMN
for representing FM-powered workflows~\cite{jasinski2022natural, vinci2024repairing}. Although effective up to the Software 2.0 era~\cite{hassan2024rethinking}, these languages are not designed to target the special characteristics of rapidly emerging and unavoidable \fms and \fm{}-based software.
This design mismatch, in turn, makes the representation of \fm{}-powered workflows using such notations infeasible.
Specifying \fm{}-powered workflows using general-purpose programming languages is also impractical, as 
these workflows have their unique semantics requiring verbose programs to handle all the edge cases or novel abstractions to represent them. 
Community-led efforts, such as the LangChain Expression Language \rev{(LCEL)}~\cite{langchainLangChainExpression} 
and GenAIScript~\cite{genaiscript}, emphasizes the need for Domain-Specific Languages (DSLs) better suited for novel semantics of \fm{}-powered workflows.

Representing \fm{}-powered workflows through such \fm{}-friendly \dsl{}s serves two purposes. First, this enables humans 
to communicate their intention when orchestrating a workflow to \fms much more clearly and at a higher level of abstraction.
Second, this increases the success rate of \fms in synthesizing workflows and automatically executing them. 
Using a \dsl
simplifies the generation of domain-specific program instances
by omitting FMs' domain-irrelevant, pre-learned knowledge.
Furthermore, using a \dsl limits the action space of \fms
by narrowing the
multitude of ways of conducting the same tasks to the most context-relevant ones~\cite{Gandhi2023}. 
As such, 
\fms are 
ensured to synthesize programs 
that adhere to domain-specific constraints in domains such as data management and computer-aided design~\cite{joel2024survey}.

% Similarly, 
Following these observations, several efforts have emerged to create \fm{}-friendly \dsl{}s
and interfaces which enhance the \fms{}'~task execution capabilities by constraining their interaction with execution environments. One example is the Office Domain-Specific Language~\cite{Gandhi2023}, a language designed to facilitate action execution within Office applications while minimizing ambiguous and overly expansive language features, thereby simplifying API invocation semantics for \fms, particularly Large Language Models (LLMs).
% LLMs. 
A parallel line of work is SWE-agent~\cite{yang2024sweagent}, which also provides a constrained interface, dubbed the agent-computer interface, that is tailored to the needs and constraints of LLMs for software engineering tasks.
Similarly, instead of having to extensively train \fms to call all available tool types, the Model Context Protocol~\citep{mcp} enables LLM-based clients to discover and execute tools at runtime via a unified interface. This enables expansion of the LLM's capabilities while reducing the likelihood of syntax errors in LLM-generated tool calls. 

Although these efforts enhance the reliability of \fms in automating processes within narrow scopes of a particular application or tool invocation, their proposed \dsl or interfaces are not applicable for using \fms to generate general-purpose workflows beyond those restricted domains. To advance beyond the current constrained application of \fms in automating processes, our overarching goal is to build on the capabilities of \fms, particularly LLMs, to reliably generate \fm{}-powered workflows that meet user-provided requirements, regardless of domain.
To achieve this goal, two main challenges must be overcome: (1) identifying the user intent and their requirements expressed in natural language and (2) then reliably synthesizing the workflows as \fm{}-friendly \dsl program instances that can be executed by language models to fulfill the identified user requirements. 

The first of these challenges, inferring user intent and refining the requirements, has already been investigated by Gallaba et al.~\cite{gallaba2025towards} in prior work.
They explored the capabilities of \fm-powered agents in refining vague user requirements across a diverse set of user scenarios. They introduced an \gallaba, named \alignmind, that is capable of refining requirements and generating detailed, step-by-step natural language instructions from an input user scenario. 
However, the subsequent challenge, reliably generating executable \dsl{} workflows from identified natural language instructions, with the help of FMs, largely remains unsolved because \fms' output often violates the specified \dsl grammar.

% On the other hand, 
Even with recent advancements, LLMs are known to exhibit a tendency for \textit{hallucination}, often producing outputs that are inaccurate, incomplete, or inconsistent with prompts~\cite{Ji2023}. Such issues are especially troubling in code generation tasks (such as our case of synthesizing \dsl workflow programs), where syntactic and semantic correctness are paramount. Therefore, providing \dsl{}s
% and interfaces 
which are specifically designed around LLM limitations (e.g., unambiguous function signatures, streamlined syntax, and comprehensive error reporting), is necessary for 
% the 
language models to operate more effectively and synthesize code that is robust or somewhat ``FM-proof''. However, even with such specialized \dsl{}s carefully curated to meet language model constraints, LLMs remain fallible and are likely to produce errors when generating DSL code~\cite{Bassamzadeh2024}. 
% \sogol{I think the following sentence can be removed as we are repeating it as the conclusion of the next paragraph in its last sentence:} In other words, guaranteeing error-free workflows, \res{which comply to the specified grammar,} through language model generation alone is unattainable, necessitating further analysis to detect and repair mistakes before execution.

Indeed, there are success stories of using language models to synthesize \dsl programs~\cite{zhang2024gh}, in which LLMs are used to directly generate GitHub Actions workflows. However, despite being \dsl program instances, GitHub Action Workflows are abundantly accessible to LLMs during their training through open-source GitHub repositories. Hence, LLMs do not face the ``low-resource language'' problem wherein the \dsl or API is novel or not widely represented in standard LLM training data. Therefore, in that instance, LLMs are successful in 
generating error-free GitHub Action Workflows without the need for implementing post-generation strategies to mitigate possible errors. On the other hand, synthesizing general-purpose workflows as program instances of low-resource, custom \dsl{}s has not been effective in practice.
In spite of reducing the attack surface of possible program synthesis mistakes through
simplifying language constructs, 
it 
does not eradicate all defects in \workflow{}s. 
Errors remain frequent, or even increase, in synthesized \dsl workflows because the LLM is effectively ``guessing'' the correct syntax or semantics without a substantial training foundation.
The desire for robust, secure, and maintainable systems underscores the need for complementary measures, e.g., detection via static analysis and automated repairs, to achieve truly reliable generation of \fm{}-powered workflows.

In this paper, we propose a novel framework that capitalizes on static analysis to detect and repair defects in \workflow{}s to ensure reliable execution in enterprise settings. Although prior research~\cite{Pan2024, wang2024large} examined bugs introduced by \fms during code generation and code translation, the defects generated by \fms during Natural-Language-to-\dsl (NL-2-\dsl) translation remain unstudied.
Therefore, we begin by creating \rev{an initial} %the first 
taxonomy of defect types in \workflow{}s, revealing \rev{$20$} %$18$ 
distinct categories of mistakes that typically appear. 
We then demonstrate how nine of these categories can be pragmatically detected via static analysis without running the workflows. To facilitate this endeavor, we build a bespoke static analyzer tailored to \workflow{}s, an area that has yet to be thoroughly explored. Lastly, our approach leverages \fms once more
to repair these defect types by incorporating direct feedback from the implemented static analyses.

Specifically, this paper investigates the following research questions: \textbf{(RQ1)} What are the characteristics of defects in FM-generated DSL workflows? \textbf{(RQ2)} How accurately can these defects be detected using static analysis? \textbf{(RQ3)} How effective are FMs at repairing these defects when provided with static analysis feedback? Our contributions can be summarized as follows.

\begin{itemize}
    \item We introduce \rev{an initial} %the first 
    taxonomy of \rev{$20$} distinct defect types commonly arising in \workflow{}s.
    \item We develop a specialized static analysis tool, \staticanalyzer, capable of detecting nine of those defect types.
    \item By implementing our repair tool for \workflow{}s, \correctionmodule, we demonstrate that these defect types can be automatically repaired by invoking \fms with actionable feedback derived from the static analysis reports.
    \item Analyzed \workflow{}s, ground-truth defect annotations, analysis results, and the prompts used during experiments are made available online for replicability.~\footnote{\url{https://doi.org/10.6084/m9.figshare.28600052}}%~\footnote{\url{https://figshare.com/s/93843302846e2827516b}}
\end{itemize}

The remainder of this paper is organized as follows.  
Section~\ref{sec:background} describes the background on executable Workflow generation with FMs. Section~\ref{sec: empirical-study} describes the design and results of our empirical study on defects in FM-generated workflows. 
Section~\ref{sec: defect-detection-repair} describes the process for workflow defect detection and repair.
Section~\ref{sec:discussion} discusses insights, limitations, and future directions. 
Section~\ref{sec:threats} describes threats to the validity of our study.
Section~\ref{sec:related_work} provides an overview of \fms and their role in automated workflow generation.
Finally, Section~\ref{sec:conclusion} concludes this work.

\section{Executable Workflow Generation with \fms }
\label{sec:background}
Going from natural language requirements to executable workflows in an enterprise setting is a multi-step process.
In this section of the paper, we define our general-purpose workflow \dsl grammar, describe a computational environment for executing these workflows, and outline our pipeline for generating the workflows in the defined \dsl using \fms, such that these workflows can be parsed and executed in the introduced execution environment.
We conclude the section by explaining a challenge in \fm{}-powered workflow generation, which motivates our investigation into detecting and fixing the defects.

\subsection{Workflow \dsl}

Based on our prior experience in modeling processes of small and medium-sized enterprises, we specify a \dsl grammar such that it can represent a varying range of these processes as workflows, from ones involving a single task to workflows consisting of multiple steps with complex control flow. We refer to this language as \wfdsl in the remainder of this paper.
While the primary objective of the \wfdsl is to represent FM-based workflows, it is an extensible and flexible grammar that adapts to any domain, whether generic or with a specific set of requirements and characteristics, such as the medical domain.

A synthesized workflow adhering to \wfdsl grammar consists of \textit{inputs field}, \textit{outputs field}, and a list of \textit{task nodes}. The inputs field captures 
global variables and user-set parameters, including their type information.
Similarly, the outputs field pertains to the expected outputs of the workflow after execution, along with their types. The task nodes correspond to the smaller, atomic steps of the workflow. We define \wfdsl as a procedural language with a diverse set of control-flow constructs consisting of \textit{regular tasks}, \textit{conditional branches}, \textit{switch cases}, and \textit{loops}. A regular task node is the equivalent of a statement in a general-purpose programming language. Similarly, conditional branch, switch case, and loop nodes correspond to if-else, switch, and for-loop syntax in a general-purpose programming language. We specify \wfdsl grammar as a TypeScript type definition in a structured, strict JSON format.

\subsection{Workflow Execution Environment}
\label{sec:execution-platforms}

The general-purpose workflow execution environment that we consider in this work is capable of executing \wfdsl programs
by orchestrating \textit{skills}, which are the most granular unit of execution in the workflow.
The execution environment can invoke four types of skills.
\textit{Native skills} pertain to using local tools, sending requests to both local and external API endpoints, and executing code snippets. \textit{Semantic skills} pertain to prompting external or internal instances of \fms, such as sending requests to \openai{}'s chat completion endpoint. \textit{Model skills} are used for invoking any machine learning model other than \fms, such as a model that was trained for sentiment analysis.
Finally, \textit{workflow skills} are used to call other \wfdsl instances, which can orchestrate the skills of all other types. 

% \subsection{Converting natural language to workflows}
\subsection{Natural-Language-to-\dsl pipeline}
\label{sec:nl2wf-converter}

\begin{figure}[t]
    \centering
     
    \includegraphics[clip, trim=0.5cm 0 0.5cm 0, width=\columnwidth]{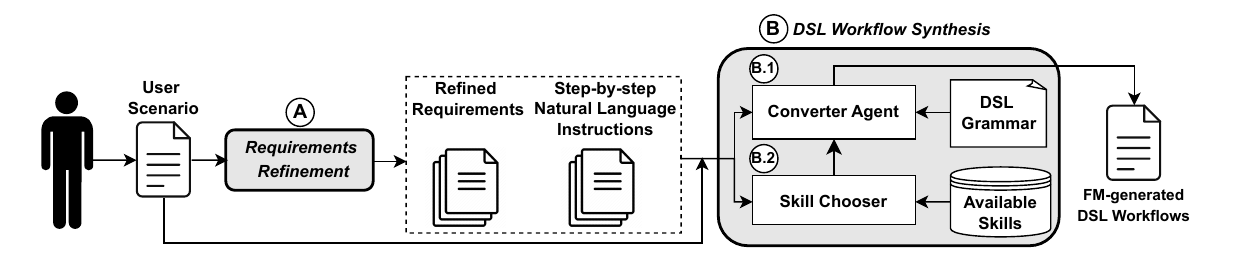}
    \caption{Overview of workflow generation with \fms using NL-2-\dsl pipeline}
    \label{fig:nlt2wf}
\end{figure}

Given a \dsl grammar and the discussed execution environment with diverse capabilities, for a specific user scenario, manually writing a detailed workflow in the defined \dsl grammar that can invoke the appropriate skill at each smaller, atomic step is a laborious endeavour.
Therefore, we implement a Natural-Language-to-\dsl (NL-2-\dsl) pipeline for automatically translating the natural language user scenario
to an executable workflow, as a \dsl program instance, with the assistance of \fms.
% \keheliya{Separate this into two phases (In figure and in text): Requirements refinement and Workflow synthesis. 
% Briefly mention that Requirements refinement can be done in practice using a multi-agent conversational system. Say, in this particular embodiment, we use AgentMind for this purpose. direct the reader to the other paper for more details on this. Then dive deep into the workflow synthesis part, as it is the focus of this paper. continue describing A and B as components of the synthesis phase. }\sogol{Done! Please double check!}
\autoref{fig:nlt2wf} illustrates the key steps involved in the process, which consists of two main parts: \textit{Requirements Refinement}~\bluecircled{A} and \textit{\dsl Workflow Synthesis}~\redcircled{B}.
We discuss these steps in detail as follows.

\bighead{Requirements Refinement~\bluecircled{A}}
In practice, it’s rare for people to outline the requirements they want to automate as a workflow in sufficient detail on their initial attempt~\cite{Wu2025,mu2024clarifygpt}. Hence, the NL-2-\dsl pipeline should first refine the vague requirements provided by users. For this purpose, a \fm{}-powered, multi-agent, conversational system can be used. In this particular embodiment, we use \alignmind~\cite{gallaba2025towards} to interact with users to refine their initial set of requirements and to generate a list of step-by-step natural language instructions corresponding to their intent, which they describe as a user scenario. 
\bighead{\dsl Workflow synthesis~\redcircled{B}}
To synthesize \dsl workflows, two main components interact with each other: the Converter Agent~\mauvecircled{B.1} and the Skill Chooser~\mauvecircled{B.2}.
% \bighead{Converter Agent~\bluecircled{A}} 
The Converter Agent~\mauvecircled{B.1} receives 
the user scenario provided by the user, the refined description of the requirements, and the corresponding natural language instruction steps, generated from step~\bluecircled{A}, as inputs.
Including the natural language instructions as an input to~\mauvecircled{B.1}, helps the workflow synthesis in two ways. First, it guides the converter agent in synthesizing the workflow, specifically including the actionable steps necessary to achieve the input user scenario.
Second, it helps validate the correctness and completeness of the generated \dsl workflow with respect to the refined requirements.
Given these inputs, the \fm in this component is prompted to generate the corresponding workflow as a valid JSON object that adheres to the \wfdsl grammar. Specifically, we prompt the \fm to generate the task nodes of the workflow in the same order as the steps outlined in the natural-language instructions, while preserving their control-flow constructs. The steps in the natural language instructions can be categorized as \textit{declarations}, \textit{conditions}, \textit{case distinctions}, and \textit{iterations}. We map each step into \wfdsl{}'s corresponding task nodes using the same control-flow construct.
% defined in the \dsl.
The declarations in the natural language instructions are mapped to regular task nodes in \wfdsl.
% workflow. 
Conditional statements in natural language are mapped to conditional branch nodes, while case distinctions map to switch case nodes. Finally, iterative instructions correspond to loop nodes in \wfdsl.

% \bighead{Skill Chooser~\redcircled{B}} 
To generate the task nodes in the workflow as proper, executable counterparts to the steps of natural language instructions,
each task node should invoke a skill from the execution environment's skills database,
matching the action described in its corresponding natural language instruction step, along with correct input/output parameters. Skill Chooser~\mauvecircled{B.2} enables this one-to-one mapping during the workflow generation. In the Skill Chooser component, we augment the prompt to the Converter Agent with the list of skills available in the skills database as context.
For each skill, we provide the skill name, accepted input fields with their types, expected output format, and a natural-language description of the skill. There are three modes available to construct the list of skills.
% There were three options available here to provide the list of skills.
In the first mode, the \fm is provided with all available skills, along with input/output as additional context in the prompt. 
The second mode queries the database for exact matches of natural-language instruction steps among available skills to include in the prompt.
In the last mode, skills with the highest semantic similarity to steps in the natural language instructions are appended as additional context.
To mitigate \fms{}' known limitation of getting distracted by irrelevant context~\cite{shi2023}, after preliminary evaluations, we chose to use the third mode.

\subsection{Motivating Example}
\label{sec: motivating-example}

% \begin{figure}[t]
%     \centering
%     \footnotesize
%     % \begin{minted}[highlightlines={3},highlightcolor=yellow]{ruby}
%     \begin{lstlisting} [language=wf, escapechar=!, caption={A defective FM-generated DSL workflow as Motivating Example}, label=motivating-example]
%   workflow: {
%     inputs: ["x>20 && y<3"]
%     tasks: [
%       { id: "task1",
%         if: {
%           condition: "x > 10 && y < 5",
%           then: "p=xy",
%           else: "q=x+y"}},
%       { id: "task2",
%         if: {
%           condition: "x < 15 && y > 0",
%           then: "s=p/2",
%           else: "t=2q"}}]}

% \end{lstlisting}  
% \end{figure}

\begin{figure*}[t]
    \centering
    \includegraphics[width=\linewidth]{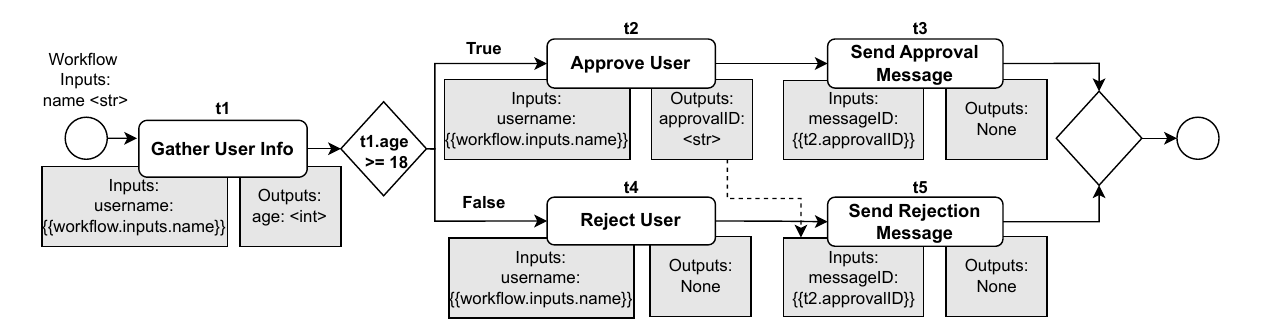}
    \captionsetup{justification=centering}
    \caption{An example of a defective workflow generated by an FM. Task t2 and Task t5, which are on different branches, have an unsatisfiable data dependency between them.}
    % \keheliya{increase the text of legend}}
    \label{motivating-example}
\end{figure*}

Similar to observations from prior work~\cite{Bassamzadeh2024}, despite our best efforts in generating valid workflow programs adhering to \wfdsl{}'s grammar, on several occasions, we observe
that generated workflows suffer from syntactic and semantic defect incidences. \autoref{motivating-example} demonstrates a simplified version of one such example we have observed in practice. 

In this example workflow, the requirement is to check if a user is old enough to be approved. If the user is 18 years or older, their profile is approved, and an approval message is sent. Otherwise, the request is rejected, and a rejection message is sent to the user. Although the generated workflow is correct at a high level, the FM introduces a subtle bug during workflow generation as follows. \textit{Task t5} (Send Rejection Message step) accepts a \texttt{messageID} as input. The output of \textit{Task t2}, \texttt{ApprovalID}, is assigned to this \texttt{messageID} field, creating an unsatisfiable data dependency because these two nodes are on mutually exclusive branches. We hypothesize that this error occurs in the \fm{}-generated workflow because the tokens in the \dsl JSON representation are generated sequentially by the \fm. Due to the limitations of autoregressive generation, the \fm cannot reason about data dependency across nodes and assumes that the output of the \textit{t2} node is visible to the \textit{t5} node. 
Although this is one example, there are many such defect patterns that require deeper reasoning on control flow and data flow to identify. 
This problem is further exacerbated by the fact that, unlike general-purpose programming languages, which \fms are well-trained on, \dsl{}s are relatively unfamiliar to them.

\section{Empirical Study of Defects in \workflow{}s}
\label{sec: empirical-study}

Ideally, for a given user scenario, after all the requirements are elicited and step-by-step instructions are available in natural language, an \fm-based agent should be able to synthesize a syntactically and semantically valid workflow in the specified \dsl.
However, as shown in Section~\ref{sec: motivating-example}, defects in \workflow{}s can affect robustness and reliability. 
Before we can effectively address this issue, we must first identify the types of defects that are occurring in practice and assess the prevalence of each defect type.
To the best of our knowledge, there exists no prior study exploring the ability of \fms in \dsl program
synthesis. 
The only similar study is the research conducted by Pan et al.~\cite{Pan2024}, which presents a taxonomy of defects when \fms are used for code translation between pairs of general-purpose programming languages, including C, C++, Go, Java, and Python.
That study mainly differs from the problem we are tackling, which involves translating natural language into workflows as program instances of a \dsl (i.e., NL-2-\dsl).
Therefore, we conduct an empirical study on plausible defects in \workflow{}s as this indicates \fm capabilities in synthesizing error-free \dsl programs. We present the results of the study as \rev{initial} %the first-ever comprehensive 
taxonomy of observed defects in \workflow{}s.

\begin{figure}[!t]
    \centering
    \includegraphics[clip, trim=0.5cm 0 0.5cm 0, width=\linewidth]{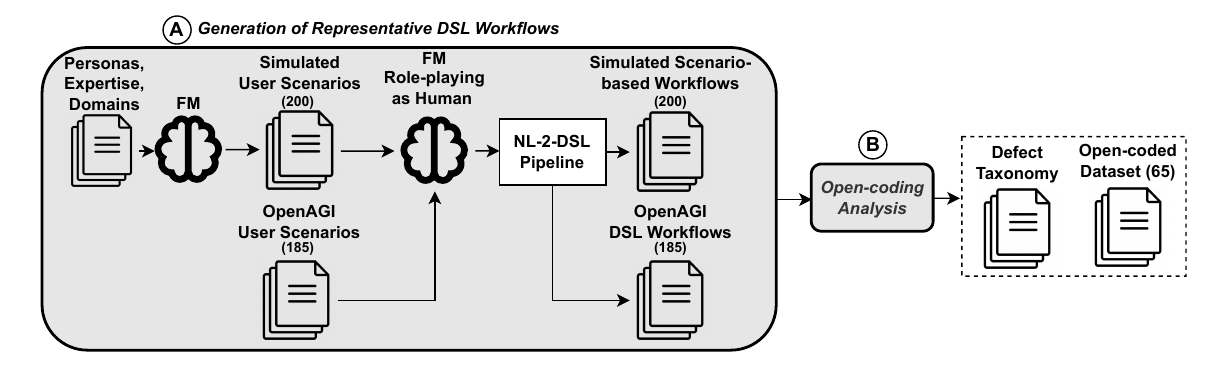}
    \caption{Overview depicting the empirical study of defects in \workflow{}s}
    \label{fig:empirical-study}
\end{figure}

\subsection{Empirical Study Design}
\label{sec: empirical-study-methodology}
\autoref{fig:empirical-study} illustrates the key steps of our empirical study, which consists of \textit{Generation of Representative \dsl Workflows}~\bluecircled{A} and \textit{Open-coding Analysis}~\greencircled{B}.
% \textit{User \res{Requirement} Elicitor \res{and Refiner}}~\bluecircled{A},
% including \textit{user-scenario generator}~\skybluecircled{A.1} and \textit{requirements generator}~\skybluecircled{A.2}, 
% \textit{Workflow Generator}~\redcircled{B}, and \textit{Open-coding Analysis}~\greencircled{C}. 
We discuss each step in detail as follows.
\bighead{Generation of Representative \dsl Workflows~\bluecircled{A}} We use our NL-2-\dsl pipeline discussed in Section~\ref{sec:nl2wf-converter} to generate the dataset of \dsl workflows for our empirical study. To effectively identify all defect types that are likely to arise in real-world situations, it is essential that the inputs of the NL-2-\dsl pipeline, namely, the initial set of user scenarios, corresponding refined requirements, and natural language instruction steps, are comprehensive and diverse. However, due to data security and privacy regulations, the real-world artifacts obtained from \rev{our} customers cannot be directly used for experimentation. As an alternative, we \rev{create a set of} %generate 
scenarios and simulate user conversations based on these scenarios to refine their corresponding requirements and generate instruction steps with the assistance of FMs. \rev{We create the set of scenarios in two steps. In the first step,} %For this purpose, 
we follow the approach described by Gallaba et al.~\cite{gallaba2025towards} to generate a list of 
% \rev{shadow} 
user scenarios %that closely approximate the authentic ones. 
\rev{that we refer to as \textit{simulated scenarios} as they closely approximate and follow the characteristics of the authentic customer scenarios.}

% As the first step of this process, 
\rev{To generate the simulated scenarios,} we prompt a \fm to generate $200$ user scenarios corresponding to a diverse set of user \textit{personas} (specifically \textit{cordial}, \textit{neutral}, \textit{rude}, and \textit{indecisive}), their \textit{level of expertise} (namely \textit{novice}, \textit{intermediate}, and \textit{expert}), and scenario \textit{domains} (including 
\textit{data analytics}, \textit{automation}, \textit{project management}, \textit{customer engagement}, \textit{customer support}, \textit{process optimization}, and \textit{automated marketing}). As an example, one user scenario generated from this process could be: \texttt{``As an experienced data scientist, I would like to receive news about AI in my inbox every morning''}. 

\rev{In the second step of creating the set of scenarios, to overcome the limitation of relying solely on synthetically generated user scenarios, we include a series of real-world, general-purpose scenarios. We choose these real-world user scenarios from the \openagi dataset~\cite{openagi}. The \openagi dataset contains $185$ natural-language scenario descriptions that are queried by users to generate real-world workflow processes. The dataset consists of 20 varying domains ranging from ``academics'', to ``mathematics'', to ``technical support for computer issues'', and ``design'', demonstrating the representativeness of the dataset across different areas. For instance, a sample \openagi scenario, selected from the academics domain, is as follows:  \texttt{``Find papers on the effectiveness of online learning platforms for language acquisition''}.}
% \keheliya{Do we need to explain this again? Can we omit the things that we already explained in Section 2.3? ->}\sogol{Edited, double check please.}

\rev{In the next step,} we 
% then 
employ the \fm to role-play as a human user and converse with \alignmind to elicit and refine the requirements for our \rev{set of simulated and \openagi} %list of synthetic 
scenarios.
% generated in the previous step.
% \res{We then use \alignmind, the interactive, multi-agent framework of Gallaba et al.~\cite{gallaba2025towards} to elicit requirements and refine them for our set of user scenarios. To do so, we employ \gpt{}-4o-mini role-playing as a human user, to express their intent for automating a tasked process in the form of a scenario, and to converse with \alignmind, intervening to refine the initial, incomplete requirements of the user. Afterwards, \alignmind generates a detailed step-by-step natural language plan from the refined requirements.}
We use the approach from Gallaba et al.~\cite{gallaba2025towards} following best practices~\cite{wang2024sotopia, abbasiantaeb2024let} to simulate the interactions between the \fm role-playing as the user and \alignmind, which is context-aware and semantically conditioned, closely mirroring authentic human-\fm dialogues. As such, we can ensure the pairs of scenarios-requirements are clear, cohesive, and correct, which in turn, align their corresponding \workflow{}s with their real-world counterpart processes.
Finally, we use our set of representative %, simulated 
artifacts as the input to the NL-2-\dsl pipeline and obtain a dataset of \rev{$385$} diverse \workflow{}s \rev{($200$ workflows corresponding to the simulated scenarios, from here on referred to as \textit{simulated scenario-based workflows}, and $185$ workflows corresponding to the \openagi user scenarios, from here on referred to as \textit{\openagi \workflow{}s})} to be further analyzed for their structural characteristics and possible defects. \openai{}’s \gpt{}-4o-mini-2024-07-18 model, with temperature set to $0.7$ and $128,000$ tokens as the context window, is used in this particular experiment.

\bighead{Open-coding Analysis~\greencircled{B}}
\label{sec:open-coding}
For creating the taxonomy of defect categories in \workflow{}s, we conduct an \opencoding analysis on 
the dataset of \dsl workflows we constructed in~\bluecircled{A}.

\rev{We start the analysis by \opencoding the set of simulated scenario-based workflows} %The analysis is carried out 
in three steps\rev{, carried out} by one researcher and one industry engineer (i.e., two inspectors). In the first step, ten workflows are randomly sampled from the initial population of 200 scenario-based simulated workflows, with stratification used to ensure that each persona, expertise level, and domain specification is selected at least once. \rev{This selected initial seed does not aim to provide adequate representation across different characteristics of the population. Rather, the goal is to calibrate inspectors' perspectives on the various defect types that may occur in \workflow{}s.}
% at least one workflow corresponding to each combination of personas, expertise, and domain specification during user-scenario generation is selected. 
\rev{For this purpose,} at the second step, each inspector independently analyzes the selected 
ten workflows and records any potential errors, their types, and their exact task node location within the workflows. To do this, they compare each workflow against its corresponding user scenario, the refined set of requirements, and the natural language instructions to identify, based on their expertise, any discrepancies and errors. 
% Afterwards, 
\rev{Inter-rater agreement reliability corresponding to the annotation of these ten \dsl workflows is calculated as a measure of \textit{Cohen's kappa} coefficient~\cite{cohen1960kappa}. Cohen's kappa quantifies the agreement between two inspectors by comparing the number of annotations both inspectors agreed on against chance-expected agreements, with values ranging from -1 (i.e., absolute disagreement) to 1 (i.e., perfect agreement), and 0 indicating agreement expected by random chance. Calculated Cohen's kappa is $0.55$, indicating moderate agreement between the two inspectors~\cite{mchugh2012interrater}.}
To \rev{resolve the discrepancies and} alleviate potential examiner bias, the inspectors meet to discuss %any 
disagreements.
% about defect incidences and to reach 
% a 
% consensus. 
% For this purpose, they compare their analysis to unify records of similar defects under the same category. If a defect incident can be categorized under an already existing defect type, that incident is added to the prevalence of that category. Otherwise, a new defect category is created to extend the taxonomy and relabel prior affected incidences. 
Identified defects by both inspectors are compared, and if they disagree on the presence or type of a defect, they discuss the discrepancy until an agreement is reached. By conducting this meeting, inspector perceptions of possible defect types align, ensuring a consolidated initial taxonomy as the baseline for continuing the next steps of the analysis. 

At the third step, inspectors continue the resampling of the \rev{simulated scenario-based workflows}, %\workflow{}s, 
adding to the set of ten initially labeled workflows, and analyze each newly sampled workflow for its defect incidences independently. After labeling each workflow, the inspectors compare their analyses to record the updated prevalence for defect categories and to extend the taxonomy when a new defect category emerges.
% necessary. 
% If any discrepancy arises, a third inspector acts as a judge to resolve the disagreement, making the final decision on the existence and type of defect under scrutiny. 
Sampling of \rev{simulated scenario-based workflows}, %\dsl workflows, %\workflow{}s, 
and labeling them, continues until no new defect types are identified. Beginning with the $50$\textsuperscript{th} sampled workflow, %\workflow{}, 
we observe no new defect types, reaching a saturation point for \rev{$15$} %the 
emergent defect categories at \totalsize open-coded \rev{simulated scenario-based workflows}. %\workflow{}s. 
\rev{Cohen's kappa across all $55$ \opencoded workflows is $0.89$, indicating almost perfect agreement between the two inspectors~\cite{mchugh2012interrater}.}
% \workflow{}s. 

\rev{To ensure the representativeness of the defect categories that emerge through \opencoding the simulated scenario-based workflows, we extend the analysis by \opencoding the \openagi \workflow{}s, following the same process described in the third labeling step. For this purpose, we randomly select two scenarios from each of \openagi{}'s ``academics'', ``creation'', ``gaming'', ``recommendation'', and ``cooking'' domains, and label their corresponding workflows (i.e., a total of ten \openagi workflows) for defect incidences. During this analysis, we record the prevalence of previously identified defect categories and investigate whether any new defect types appear in the \openagi workflows. Open-coding the \openagi workflows does not reveal any new defect category beyond those already identified in simulated scenario-based workflows.}
% \workflow{}s.}
As such, a taxonomy of \rev{$15$} defects is created by combining the individual taxonomies developed by the two inspectors \rev{through \opencoding a total of $65$ \workflow{}s}. 

% (2) In the second step, the inspectors analyze the remaining $45$ workflows for their incidences of defects. (3) In the last step, the inspectors compare their analyses to record the updated prevalence for defect categories and to extend the taxonomy when necessary. If any discrepancy arises, a third inspector acts as a judge to rule out the disagreement, making the final decision on the existence and the type of the defect incident under scrutiny. As such, a taxonomy of $13$
% \keheliya{how many?} 
% defects is created by combining the individual taxonomies created by the two inspectors. 
To ensure the completeness of the taxonomy, based on relevant literature on general-purpose programming languages~\citep{xuan2016nopol, srivastava2010program, sun2022consistency}, \rev{we also include three additional defect types (i.e., violation of concurrency constraints, \badjson, and \badexp). This extension is motivated by well-known struggles of \fm{}s in generating structured outputs such as \json objects~\cite{shorten2024structuredrag} and their limitations in generating correct mathematical expressions~\cite{li-etal-2024-evaluating-mathematical}.} \rev{Additionally, based on the observed behavior of the synthesizer \fm in manipulating workflow logic (i.e., omission of logical steps) and generating faulty outputs, we extend the taxonomy to account for the possibility of \fm{}s introducing incorrect logic or incorrect inputs in generated workflows.}
After this process, we obtain a final taxonomy of \rev{$20$} %$18$
potential defect types in \workflow{}s.

\subsection{Taxonomy of Defect Incidences in \workflow{}s}
\label{sec: taxonomy}

After conducting the empirical study in the previous section, we are able to address the first research question regarding defects in \workflow{}s.

% We introduced the first-ever comprehensive taxonomy of defect incidences in \workflow{}s as a result of assessing the abilities of \fms in synthesizing \dsl programs. We discuss our taxonomy from two aspects, one discovering plausible defect types tainting these workflows and then analyzing the prevalence of each of these defect categories across the studied data. For this purpose, we address the following research question.

% Based on

\smallskip
\noindent\textbf{\reqone}

We characterize the discovered defects in \workflow{}s by categorizing them into different types and calculating the prevalence of incidences for each defect type.
% \keheliya{Separate the main RQ and sub RQs in the macro. Have some connecting sentences to bridge the main RQ to the sub-RQS.}

% We categorize defect incidences observed while conducting \opencoding analysis on the curated dataset of \workflow{}s and calculate the prevalence for each of the emerged defect types.

\begin{figure*}[t!]
    \centering
    \includegraphics[width=\linewidth]{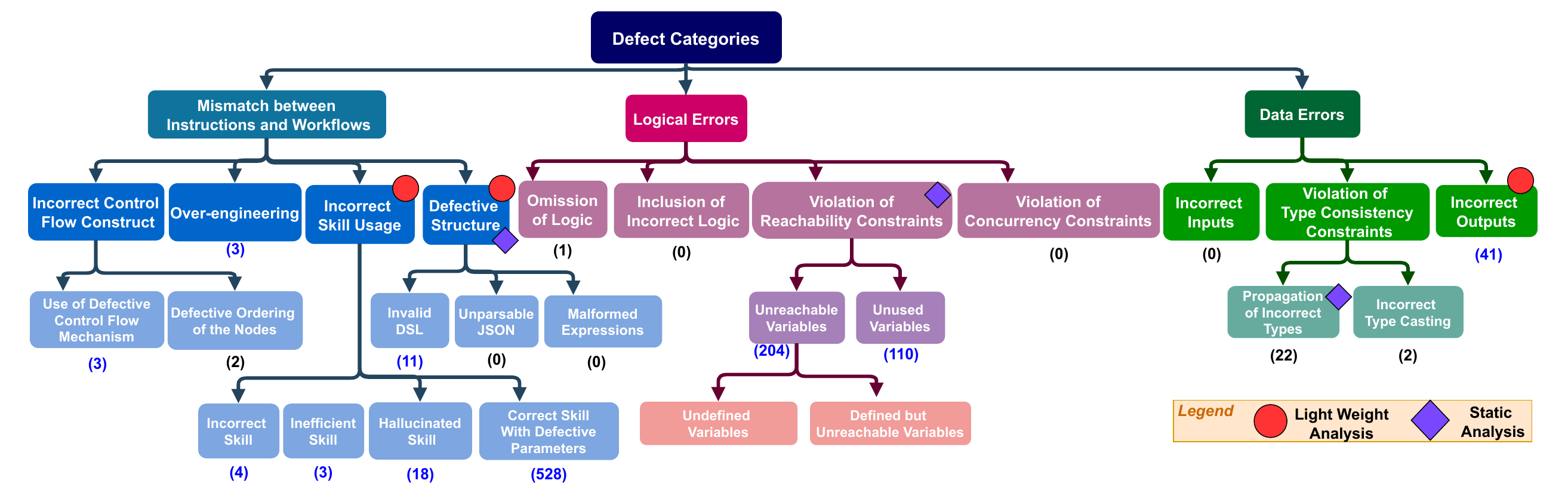}
    \captionsetup{justification=centering}
    \caption{The taxonomy of \rev{$20$} defect types in FM-generated DSL workflows, organized into three main categories. Numbers in brackets indicate the total count of each defect found in our \rev{65}-sample open-coded dataset.}
    % \keheliya{change control flow defects: sub node. Rename "Removal of logic" to "Omission of logic"}}
    % \keheliya{increase the text of legend}}
    \label{fig:taxonomy}
\end{figure*}

\bighead{Defect categories}
% (\rqoneone)}
\label{sec:defect-categories}
\autoref{fig:taxonomy} illustrates the taxonomy of defect categories in \workflow{}s. There are \rev{\defectcount} distinct types organized into three distinct groups:
% categories: 
\textit{Data Errors}, \textit{Logical Errors}, and \textit{Mismatch between Requirements and Workflows}.

\head{Data errors:}
Similar to general-purpose programming languages~\citep{Pan2024}, workflows involve data manipulation and propagation. Thus, many defect incidences in these \workflow{}s are also caused by data-related issues. 

\begin{description}
    
\item[Incorrect Inputs in Workflows:]
% This means 
The \fm has hallucinated and has assigned faulty values to the inputs field of the \workflow{}. 
% the \workflow{} includes a faulty inputs field---the \fm has hallucinated while assigning the inputs field of the workflow. 
For instance, \fm has removed a user-specified parameter or has added an incorrect global variable in the inputs. 

\item[Incorrect Outputs in Workflows:]
The \fm has not adhered to the specific user instructions for structuring the outputs field of the \workflow{}.
% The \workflow{} includes a faulty outputs field---the \fm has not adhered to the specific user instructions for structuring the workflow's outputs. 
For instance, while the user instructions require the outputs field to include only the output parameters of the last task node of the workflow, 
the outputs field includes the output parameters of some other node in the workflow.  

\item[Violation of Type Consistency Constraints:]
The inputs to the workflow itself (global inputs) can be assigned as inputs to any one of the task nodes.
The output of any one of the preceding task nodes can also be assigned as inputs to the task nodes.
If in any of these situations, an assignment leads to an incompatibility of types, we label it as a type consistency constraint violation.
Situations where type casting can cause loss of precision (e.g., type casting from float to integer) are also labeled as defective.

\end{description}

\head{Logical errors:}
These defects pertain to the logical mismatch between the user requirements and their corresponding \workflow{}s. 

\begin{description}

\item[Violation of Reachability Constraints:]
The \workflow{}s could include erroneous data dependencies, which can be classified
% We categorize these 
into two categories: (1) unreachable variables, i.e., variables that are used as inputs in a task node, but those variables do not have any values by the time that task node is executed, and
(2) unused variables, i.e., variables produced as the outputs of one of the task nodes but never used anywhere else in the workflow.

% the \workflow{} includes \undefvar{}s, which pertain to use cases of either completely undefined variables or use cases of variables which are previously set but are defined in branches with circular dependency on the target use case. Second, the \workflow{} set variables but it never uses them in proceeding task nodes.

\item[Omission of Logic:]
The \workflow{} omits one or more steps of the corresponding natural language instructions.
% requirements. 

\item[Inclusion of Incorrect Logic]
The \workflow{} includes an additional incorrect step compared to the corresponding natural language instructions. 
 
\item[Violation of Concurrency Constraints:]

The \workflow{} violates resource constraints and bound usage of its execution specifications. For instance, two task nodes in different branches are trying to write to the same variable concurrently.

\end{description}

\head{Mismatch between instructions \& workflows:}
This category pertains to defect incidences caused by semantic, syntactic, and structural differences between the natural language instructions and their corresponding \workflow{}s. 

\begin{description}

\item[Incorrect Control Flow Construct:]
The \workflow{} contains erroneous control-flow dependencies. These errors can occur due to two distinct situations. First, the \workflow{} uses a different control flow construct compared to the corresponding step mentioned in the natural language instructions. Second, the \workflow{} task nodes are ordered differently from the steps described within the corresponding natural language instructions.
% , leading to erroneous data and control flow within the workflow.

\item[Over-engineering:]
The \workflow{} includes additional nodes compared to the original natural-language instructions, which are inherently correct but not required in practice. These over-engineered nodes may include unnecessary validation steps or intricate analysis to satisfy rare edge cases.

% For instance, assume the target requirement of a workflow involves two sequential steps 
% \textit{Randomly\_Choose\_a\_Letter} and \textit{$Is\_Letter\_Lowercase$}. The letter outputted from the first step is used by the second step to investigate whether it is in lowercase or uppercase. In practice, the former step always returns a letter, regardless of whether it is in lowercase or uppercase. Hence, checking it for not being \textit{None} is unnecessary. However, while generating the corresponding workflow, the \fm adds an additional task node between corresponding nodes to \textit{Randomly\_Choose\_a\_Letter} and \textit{$Is\_Letter\_Lowercase$}, ensuring the output letter of the first task is not \textit{None}. This additional step is inherently correct but unnecessary, pertaining that the \fm has ``over-engineered'' the generated workflow. 

\item[Defective Structure:]
The \workflow{} is structured incorrectly. These errors occur due to three reasons. First, the \workflow{} is structured as an unparsable \json object. Second, the workflow violates the specified \dsl{} grammar. And lastly, the \workflow{} includes malformed conditional or switch case expressions.

\item[Incorrect Skill Usage:]
The \workflow{} implements a step specified in the natural-language instructions using an incorrect skill, manifesting as four distinct situations.
First, the workflow implements the target step of the instructions using an incorrect action from the set of available skills. Second, the \workflow{} implements the target step with a correct but suboptimal action. This means the decision of the \fm for choosing that specific skill for implementing the corresponding natural language instruction
step can be improved. Third, the \workflow{} attempts to implement the target step with a non-existing action. This means the \fm has hallucinated the skill while generating the workflow. Finally, the \workflow{} implements the target step with a correct skill, but the skill is utilized with defective input and/or output parameters within the workflow. This could mean that, when generating the workflow, the \fm has hallucinated the skill's parameters and removed some of them, for example.  

\end{description}
\begin{myframe}[width=\linewidth, top=0pt,bottom=0pt,left=0pt,right=0pt,arc=0pt,auto outer arc]
\small{\textbf{\underline{\textit{\rqoneone{}.}}} The taxonomy of defects in \workflow{}s comprises \rev{\defectcount} categories, sorted into three distinct groups.} 
\end{myframe}
\bighead{Prevalence of defects}
% (\rqonetwo)}
\label{sec:defect-prevalence}
Out of \rev{$65$} \opencoded \workflow{}s, only seven workflows are not tainted with any incidences of defects, with the remaining \rev{$89.23$\% ($58$ out of $65$)} containing at least one defect incident. On average, a workflow contains $15$ defects
% an actual average count of 
(mean=\rev{$14.65$} rounded to \rev{$15$}, accounting for the discrete nature of the prevalence of defects) 
while $24$ incidences is the maximum count of observed defects in a single workflow. The high prevalence of defects in \dsl workflows, %\workflow{}s, 
evident by these statistics, underscores the magnitude of the problem and highlights the need for effective mitigation strategies.

\autoref{fig:taxonomy} also illustrates 
the prevalence of each of the defect types (below each category and in brackets) across the \opencoded dataset.
% The prevalence of each of the defect categories across the dataset of $55$ \opencoded workflows is also depicted in \autoref{fig:taxonomy} below each category in brackets. 
% The three defect types \badjson, \baddsl, and \badoutput are counted once per each affected workflow. Meanwhile, the rest of the defects are counted calculated as the total count of incidences for a specific defect type across the dataset. 
The most frequent defect type corresponds to \badparameter with \rev{$528$} defect incidences. Unreachable and unused variables are the second and third most frequent defect types with \rev{$204$} and \rev{$110$} 
% $185$ and $101$ 
incidences, respectively. \rev{As discussed in Section~\ref{sec:open-coding}, none of the annotated workflows is tainted} %in the examined sample of workflows, 
% we did not encounter any workflow 
with unparsable \json structure, \rev{additional} incorrect logic, \badexp, %hallucinated 
\rev{incorrect} inputs field, or violation of concurrency constraints. \rev{However, for completeness, and based on prior studies, we extend our taxonomy to include these defect types as well. The prevalence of these defect categories is noted as zero in \autoref{fig:taxonomy}.} 
% However, these defect categories are added to the taxonomy as similar defect incidences are observed in the dataset. 
% For instance, ``defective outputs field'' is a frequent defect type in \workflow{}s, affecting $67.27\%$ (i.e. $37$ out of $55$) of the dataset. Hence, it is likely for \workflow{}s to also contain a ``defective inputs field'' even though no such incident is observed in the current dataset. A similar argument is used to extend the taxonomy with \badjson and \badexp. If \workflow{}s are structured as invalid \dsl{}s (i.e., three incidences), it is likely for these workflows to also suffer from invalid \json and malformed expression structures, although no such incident is observed in the current population of investigated workflows. Furthermore, if \fms can incorrectly omit logical steps of the requirements when generating workflows, they can also hallucinate and include incorrect steps during workflow generation. Finally, as discussed in \autoref{sec:open-coding}, to ensure the comprehensiveness of the taxonomy and based on literature,``violation of concurrency constraints'' is included as a defect category to account for the violation of resource constraints and bound usage at the time of execution. However, to record the prevalence of these incidences, the execution of the workflows is needed, which is out of the scope of our paper. Hence, we do not report the prevalence of resource constraint violations. 

\begin{myframe}[width=\linewidth, top=0pt,bottom=0pt,left=0pt,right=0pt,arc=0pt,auto outer arc]
\small{\textbf{\underline{\textit{\rqonetwo{}.}}} \rev{$89.23$\%} %$82.27$\% 
of \workflow{}s contain at least one defect incident, with an average of $15$ defects per workflow. Among all plausible defect types, mismatches between instructions and their corresponding generated workflows dominate defect incidences, while data errors are the least frequent defects tainting the \rev{workflows}.}
% \workflow{}s.} 
\end{myframe}

\rev{\subsection{Generalizability of the Empirical Study}}\label{sec:empirical-study-generalizability}
\rev{In our empirical study (Section~\ref{sec: empirical-study-methodology}), we only open-code workflows that are generated by prompting \openai{}'s \gpt{}-4o-mini-2024-07-18 model.
% and to repair their detected defect incidences. 
Consequently, the observed defect incidences in these \workflow{}s may not extend to workflows that are synthesized by other models, such as \meta{}'s \llama. However, recent benchmarking results have shown that model performances from all frontier labs have converged, which can be attributed to 
% the relatively similar model architectures and 
likely heavy overlaps in their pre-training data~\cite{benaich2024state}. Therefore, we hypothesize that using different \fm{}s to generate \dsl workflows would have a negligible influence on the types of defects that may occur.}

\rev{To confirm our hypothesis, we investigate whether different synthesizer \fm{}s generate workflows that are tainted with defect types that are not included in our taxonomy.
For this purpose, we randomly selected ten scenarios from the population of $200$ simulated user scenarios, alongside their elicited requirements and natural language instructions used in the empirical study (Section~\ref{sec: empirical-study-methodology}). We then employ two different workflow synthesizer \fm{}s, Meta's Llama-3-70B and Alibaba's Qwen-7B (hereafter referred to as Llama and Qwen, respectively) and follow the same procedure outlined in Section~\ref{sec: empirical-study-methodology} to generate \dsl workflows for the selected simulated scenarios and annotate them for their occurring defect types and their prevalence. We specifically choose these two models to conduct a comprehensive analysis over the plausible confounding effects of varying characteristics of synthesizer \fms (i.e., open-source or proprietary, different architectures, and number of learnable parameters) on occurring defect types in \workflow{}s.}

% \usepackage{booktabs}
% \usepackage{multirow}
% \usepackage{graphicx}
\begin{table}[!ht]
\centering
\captionsetup{justification=centering}
\caption{Prevalence of defect types across \dsl workflows generated by Llama and Qwen}
\label{tab:taxonomy-generalizability}
% \resizebox{0.6\textwidth}{!}{ %
\footnotesize
\setlength{\tabcolsep}{3pt}
\begin{tabular}{@{}lcc@{}}
\toprule
\textbf{Defect Categories} & 
\textbf{\begin{tabular}{@{}c@{}}Llama-generated\\DSL Workflows\end{tabular}} & 
\textbf{\begin{tabular}{@{}c@{}}Qwen-generated\\DSL Workflows\end{tabular}} \\
\midrule
\multicolumn{3}{c}{\textit{\textbf{Data Errors}}} \\
\midrule
Incorrect Inputs & 5 & 1 \\
Incorrect Outputs & 2 & 7 \\
Propagation of Incorrect Types & 7 & 3 \\
\midrule
\multicolumn{3}{c}{\textit{\textbf{Logical Errors}}} \\
\midrule
Unreachable Variables & 51 & 51 \\
Unused Variables & N/A & 7 \\
Inclusion of Incorrect Logic & 1 & N/A \\
Omission of Logic & 1 & N/A \\
Use of Defective Control Flow Mechanism & 4 & 1 \\
\midrule
\multicolumn{3}{c}{\textit{\textbf{Mismatch between Instructions and Workflows}}} \\
\midrule
Over-engineering & N/A & 1 \\
Incorrect Skill & 4 & 10 \\
Hallucinated Skill & N/A & 11 \\
Correct Skill with Defective Parameters & 53 & 75 \\
Invalid DSL & N/A & 1 \\
Unparsable JSON & 10 & 10 \\
\bottomrule
\end{tabular}%
% }
\end{table}

\rev{\autoref{tab:taxonomy-generalizability} presents the identified defect types and their prevalence across \dsl workflows generated by Llama and Qwen.
As can be observed from the table, defect types annotated in Llama- and Qwen-generated workflows remain similar to the taxonomy of defect types created through annotating \gpt{}-generated workflows. Only the prevalence of defects varies across workflows grouped by synthesizer \fm{}s. Thus, we can conclude that our initial hypothesis on the consistency of occurring defect types, regardless of the used synthesizer \fm, %for generating \dsl workflows, 
holds true. This, in turn, confirms the extensibility of our defect taxonomy to different synthesizer model families with varying characteristics.}

Armed with \rev{our} %this 
taxonomy of common defects, we then design and implement a toolchain to automatically detect and repair the most frequent and programmatically identifiable issues.

\section{Detection and Repair of Workflow Defects}
\label{sec: defect-detection-repair}

% In the previous section, we discuss our conducted study for pioneering the evaluation of \fm abilities in \dsl program synthesis and introduce the first-ever taxonomy of defects in \workflow{}s. 
In this section, we present \staticanalyzer, a tool for detecting defects in FM-generated workflows and \correctionmodule, a tool for repairing workflow defects with the assistance of static analysis feedback in \rev{\dsl workflows}.
% FM-\rev{generated} %synthesized 
% \dsl programs.
% the newly discovered defect types within \workflow{}s. 

\subsection{\staticanalyzer: The Static Analyzer for \workflow{}s}
\label{sec:static-analyzer}
\begin{figure}[t]
    \centering
    \includegraphics[width=0.99\linewidth]{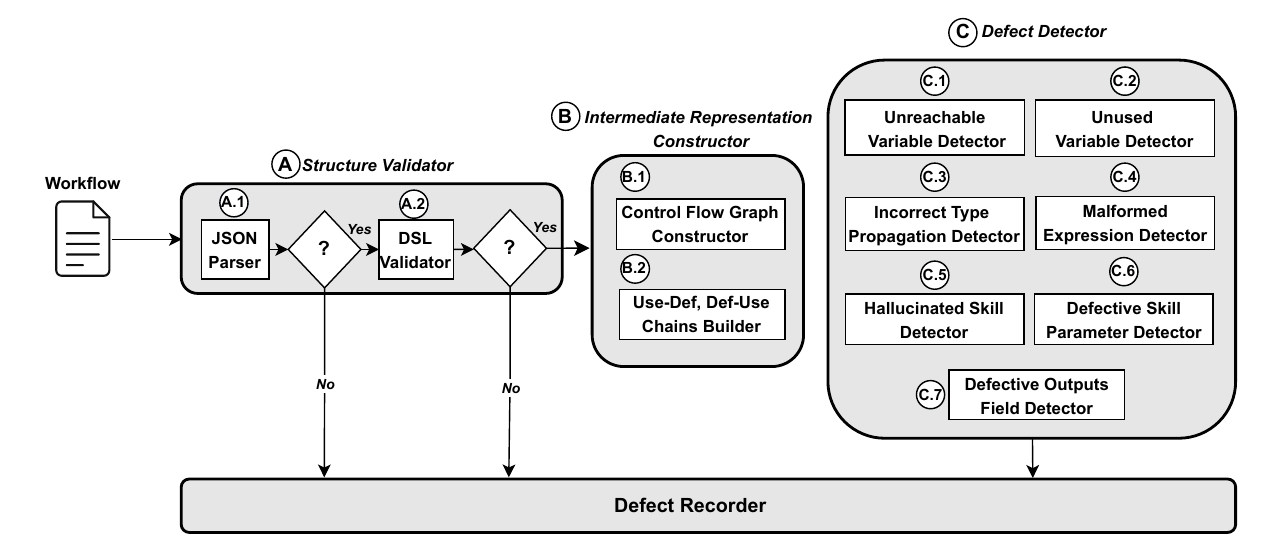}
    \caption{The overview of \staticanalyzer: The static analyzer for \dsl workflow defect detection}
    % \keheliya{There should be an arrow going from detector to defect recorder. Rename detector to defect detector. Graph constructor should be CFG constructor or Control Flow Graph Constructor}}
    \label{fig:static-analyzer-overview}
\end{figure}

We develop \staticanalyzer, as the first-of-its-kind static analyzer to automatically detect defect types frequently observed in \workflow{}s. \rev{\staticanalyzer{}'s novelty lies within the application target, i.e., \staticanalyzer is the first-ever tool that applies well-established defect detection techniques, that are designed for general-purpose programming languages, to \workflow{}s for detecting their defect incidences.} %These 
\rev{\staticanalyzer{}'s detectable defect types encompass all categories that can be identified without human intervention or workflow execution. For instance, incidences of defects such as over-engineering or inefficient skill cannot be detected without human intervention; human judgment is required to determine whether a more optimal action exists for implementing the target step within the workflow. Similarly, it is necessary to execute the workflows to detect their violations of concurrency constraints. Thus, detecting these defect types is out of the scope of \staticanalyzer.} As such, \staticanalyzer is able to detect the following \detectcount defect types from the defect taxonomy in Section~\ref{sec: taxonomy}: \badjson, \baddsl, \undefvar, \unuse, \badtype, \hallucinatedskills, \badparameter, \badexp, and \badoutput.
These correspond to defects in our taxonomy (\autoref{fig:taxonomy}) that are marked as detectable via lightweight and static analyses. \rev{The former, i.e., lightweight analysis, corresponds to detection techniques that do not rely on intermediate representations of the \workflow{}s (for more details, please refer to Intermediate Representation Constructor~\redcircled{B} below). The latter, i.e., static analysis, refers to detection techniques that operate on the intermediate representations of the workflows.}
% analysis in our taxonomy (\autoref{fig:taxonomy}).
We illustrate an overview of the tool in \autoref{fig:static-analyzer-overview}, which comprises three key steps, namely: \textit{Structure Validator}~\bluecircled{A}, \textit{Intermediate Representation Constructor}~\redcircled{B}, and \textit{Defect Detector}~\greencircled{C}. We describe each of these steps in detail below.

% \noindent\underline{\textbf{\json parser \& \dsl synthesizer.}} 
\bighead{Structure Validator~\bluecircled{A}}
First, \staticanalyzer receives a \workflow as a string input and attempts to parse it into a valid \json~\skybluecircled{A.1}. If the input \workflow is successfully parsed, then \staticanalyzer attempts to instantiate the \json object as a valid \dsl program adhering to our specified grammar~\skybluecircled{A.2}, with the support of \pydantic~\citep{pydantic} type checking library. If the \workflow cannot be parsed into a valid \json or the instantiated \dsl program does not follow the grammar, \staticanalyzer records a defect incident for \badjson or \baddsl, respectively.

% \noindent\underline{\textbf{Graph constructor.}} 
\bighead{Intermediate Representation Constructor~\redcircled{B}}
As the next step, \staticanalyzer constructs an intermediate representation from the \dsl workflow. Specifically, a control flow graph (CFG)~\citep{allen1970control} is constructed~\mauvecircled{B.1} by parsing the 
% JSON
\dsl representation of the workflow. Afterwards, from the constructed CFG %control flow graph 
\staticanalyzer builds \textit{use-def} and \textit{def-use chains}~\mauvecircled{B.2}.
% \usedefdefuse~\citep{hooda2024large, parasaram2023rete}~\mauvecircled{B.2}. 

\bighead{Defect Detector~\greencircled{C}}
Receiving the constructed intermediate representation as input, the defect detector component is executed to identify incidences of various defects. Using the constructed \usedef, \staticanalyzer tracks whether a used variable in a target node of the CFG is assigned with a value (i.e. outputted) in a prior node~\tealcircled{C.1}. Similarly, using \defuse, the tool evaluates whether each defined variable in a target node is used in at least one following node~\tealcircled{C.2}. Furthermore, using \usedefdefuse, \staticanalyzer analyzes the propagation of variables for ensuring the consistency of their primitive data types between their definition and usage~\tealcircled{C.3}. \staticanalyzer also evaluates the well-formedness of existing expressions in conditions of loops and cases of switch statements~\tealcircled{C.4}. For this purpose, the tool attempts to parse each expression into a valid \abst (AST). If the target expression is successfully parsed into an AST, reachability and usage of the expression's variables are further analyzed by adding them to \usedefdefuse.  
% \keheliya{This part is not very clear and the text is missing the links to C3 and C4?}\sogol{Some parts of the sentence have been commented out by accident. Fixed now!}
\staticanalyzer can also detect whether \taskaction{}s in \dsl workflows are hallucinated~\tealcircled{C.5}. To do this, \staticanalyzer compares each \taskaction used within the workflow against all available skills stored in the skills database of the workflow execution environment. If the target \taskaction does not exist within the database, that action is marked as hallucinated. By conducting a similar comparison, \staticanalyzer is able to identify existing \taskaction{}s that are used with incorrect input and/or output parameters within the workflow~\tealcircled{C.6}. 
% \staticanalyzer also evaluates the well-formedness of existing expressions in conditions of loops and cases of switch statements \textit{(C.6)}. For this purpose, it attempts to parse each expression into a valid \abst (AST). If the target expression is successfully parsed into an AST, reachability and usage of the expression's variables are further analyzed by adding them to \usedefdefuse. 
Finally, \staticanalyzer also detects whether the outputs field of the workflow adheres to the specific predefined format passed as user instruction at the time of workflow generation~\tealcircled{C.7}. 

\subsection{\correctionmodule: The \fm{}-based Repair Tool for \workflow{}s}
\label{sec: correction-module}
Next, we use \correctionmodule to repair defective \workflow{}s. \correctionmodule operates by prompting a \fm to repair defect incidences in an input workflow per defect type, incorporating \staticanalyzer{}'s report of detected defects as direct feedback for an accurate and effective repair.
This key idea of providing feedback from the defect detection step is inspired by prior work~\cite{chen2024teaching, gou2024critic} where the structured output from a verifier component, such as a logic checker, code interpreter, or a test suite, is used alongside the FM to increase the robustness of output generation\rev{, i.e., the application of a verifier-in-the-loop for automatically repairing defects when FMs are used as synthesizers of general-purpose programming languages}.
The intuition is that the direct, unambiguous feedback from the defect detector %detection tool 
precisely signals to the \fm where its output deviates from the desired criteria. \rev{As \staticanalyzer can only automatically detect nine defect types, from the 20 emergent defect categories in the created taxonomy, \correctionmodule can also repair the incidences of these defect types across the \workflow{}s.}
Similar to prior efforts implementing \textit{test-time iterative strategy} for improving the success rate of \fms~\cite{xia2024agentless, wang2025openhands}, \correctionmodule also iteratively prompts the \fm to repair the input workflow's defect incidences.

To ensure the effectiveness of \correctionmodule{}'s prompts, a series of prompt engineering investigations is conducted. At each step, the improvement of \correctionmodule{}'s repair performance 
is validated against a set of ten defective \workflow{}s. First, a prompt is carefully crafted for each specific defect type, tailored to repair instances of that defect category based on feedback from the static analyzer report. For each target defect type, specific additional instructions are appended to the prompt as context information. For instance, when repairing an \baddsl, the \dsl grammar is provided as a TypeScript type definition. For correcting \badoutput, the specific user instruction for formatting the outputs field is provided. To correct \hallucinatedskills and \badparameter, a list of available skills in the execution environment is appended to the prompt. At the second step, spell and grammar checkers are used to improve clarity and minimize any content ambiguities~\cite{kamath2024scope}. Next, few-shot prompting and in-context learning are applied to guide the \fm{}'s inference process~\citep{brown2020language}. Finally, \openai{}'s~\cite{openai2024prompt} and \an{}'s~\cite{anthropic2024prompt} prompting guides are consulted to refine the prompts such that their efficacy is maximized. The full prompts used by \correctionmodule are available in the replication package.

% Detected defects by \staticanalyzer in an \workflow are attempted to be repaired by \correctionmodule. \correctionmodule operates by prompting
% \openai{}’s \gpt{}-4o-2024-08-06 (with temperature set as 0.7 and 128, 000 tokens as the context window)
% using \staticanalyzer{}'s report as a ``hint'' for an accurate and effective repair. Alongside \staticanalyzer{}'s report, the defect type and the defective workflow are added as prompting context. Furthermore, based on the defect type, additional instructions are appended. For example, when repairing an \baddsl, the DSL schema is provided as a TypeScript type definition. For correcting \badoutput, the specific user instruction for formatting the outputs field is provided. For correcting \hallucinatedskills and \badparameter, a list of available skills in the execution environment is appended to the prompt instructions. The full prompts used by \correctionmodule are available in the replication package.

\subsection{Experiment Setup}
\label{sec: detection-repair-experiment-setup}

\begin{figure}[t]
    \centering
    \includegraphics[width=\linewidth]{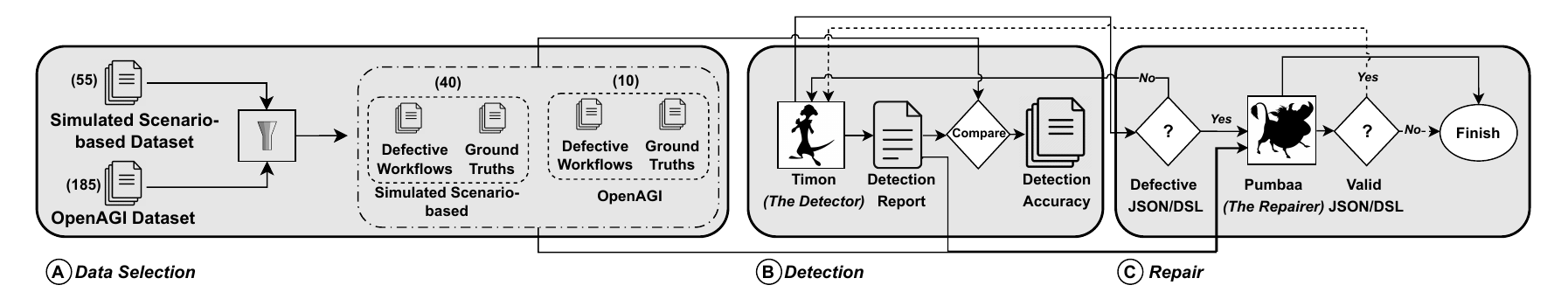}
    \caption{The experiment design}
    % \keheliya{change the WfWizer to Detector Module to be consistent with Repair Module?}}
    \label{fig:study-design}
\end{figure}

To evaluate the performance of our automatic detection and repair tools for defect incidences in \workflow{}s, we set up an experiment with three key stages, namely \textit{Data Selection}~\bluecircled{A}, \textit{Detection}~\redcircled{B}, and \textit{Repair}~\greencircled{C}, demonstrated in \autoref{fig:study-design} and discussed in detail next.   

\bighead{Data Selection~\bluecircled{A} \& Detection~\redcircled{B}}
\label{sec: study-data-selection-defect-detection}
\rev{We investigate the performances of our tools in two folds. First,} from the total population of \totalsize \rev{\opencoded simulated scenario-based workflows} %\workflow{}s 
\rev{(Section~\ref{sec:open-coding}~\redcircled{A})}, we randomly select \samplesize %of them 
\rev{and evaluate defect detection and repair performances of \staticanalyzer and \correctionmodule across them}. \rev{We also investigate the performances of \staticanalyzer and \correctionmodule on \dsl workflows that are generated from real-world user scenarios. For this purpose, we use the same ten \openagi \workflow{}s that are \opencoded for creating the defect taxonomy. For more details on \openagi user scenarios and their corresponding workflows, please refer to Section~\ref{sec: empirical-study-methodology}~\redcircled{A}}. \rev{The selected \samplesize simulated scenario-based workflows alongside the ten \openagi workflows form the evaluation set of our experiment.}
For each sample in
% of 
the evaluation set \rev{(either a simulated scenario-based or an \openagi workflow)}, \staticanalyzer is executed to record incidences of different defect types in that sample. 
We then compare the result report of \staticanalyzer for each sample against its \opencoded ground truth. %(from Section~\ref{sec: taxonomy}).
% In case of any discrepancy between the existence of a defect incident or the type of an existing defect, we would record a performance error for \staticanalyzer{}. 
By conducting this comparison, we can categorize the defect incidences into three groups. Defect instances in ground truth and also the \staticanalyzer{}'s identified set are \truepositives (\tp{}s).
Instances that only appear in the \staticanalyzer{}'s identified set are \falsepositives (\fp{}s), while instances only appearing in the ground truth are \falsenegatives (\fn{}s). 
% A detected defect incident by \staticanalyzer is a \tp, if the defect type and its locating node within the target \workflow is the same as the corresponding ground truth. If a detected defect by \staticanalyzer does not exist within the ground truth, that defect incident is categorized as \fp. Finally, if a ground truth defect is not detected by \staticanalyzer, that defect incident is marked as \fn. 
% We conduct this comparison for each sample in the evaluation set, per each defect category. 
To evaluate the performance of \staticanalyzer, we use the \tp{}s, \fp{}s, and \fn{}s and calculate the tool's \precision and \recall.
% as follows.  

% \begin{align}
% Precision= \tp{}s/(\tp{}s+\fp{}s) \\
% Recall= \tp{}s/(\tp{}s+\fn{}s)
% \end{align}

\bighead{Repair~\greencircled{C}}
\label{sec:study-defect-repair}
We start the repair of the \workflow{}s in the evaluation set with \badjson and \baddsl defect types. If the \workflow cannot be parsed into a valid \json or the generated program does not follow the specification of \wfdsl grammar, 
% TypeScript type definition grammar,
analyzing it for incidences of other defect types is futile. Hence, if \staticanalyzer detects either of these defects for the sample workflow, it pauses the detection process and sends the invalid workflow to \correctionmodule. If \correctionmodule is able to correct the input such that it can be successfully parsed as a JSON object and instantiated as a valid \dsl program, the corrected \dsl program is sent back to \staticanalyzer to continue the defect detection process for other defect categories and produce a comprehensive report for all incidences of \detectcount defect types. Otherwise, that input workflow is marked as ``incomprehensible'' and the rest of the detection and consequent repairs are skipped.

After retrieving \staticanalyzer{}'s report of defect incidences for each \workflow in the evaluation set, using \correctionmodule, we attempt to repair the sample workflow incorporating \staticanalyzer{}'s report as direct feedback. 
To demonstrate the impact of inherent capabilities of the used \fm on the repair performance of \correctionmodule, we design our experiment to include both 
% \rev{To ensure \correctionmodule's operation on its optimum capabilities, the best performing \fm for repairing defect incidences should be selected. To do so, we design our experiment to involve both 
proprietary and open-source \fms. Used models are of varying parameter counts, ranging from smaller variants to larger ones, and different architectures, both dense and mixture of experts (MoE). 
% \rev{We} investigate and compare \rev{the performance of these models in repairing defective \workflow{}s.}
% with each other, their performances in repairing the defective \workflow{}s. 
% Specifically,
In particular, we use and compare the repair performance of three different families of models including \openai{}'s \gpt{}-4o-2024-08-06 (as the proprietary model, with an estimation of $200$ billion parameters and a MoE architecture), Meta’s Llama-3-70B, and Alibaba’s Qwen-7B (both as open-source models, with dense architectures and a medium and small count of learnable parameters, respectively). From here on, we refer to these models as
% to be referred as 
\gpt-4o, \llama{}, and \qwen,
% here on,
respectively. To minimize any confounding factors which may influence the performance of these models, we set their decoding parameters 
% of them 
to be similar, with their temperature set as $0.7$ and their context window limit set as the maximum which consists of $128, 000$ tokens for \gpt{}-4o and $8,192$ tokens for both \llama and \qwen models. 

For each \fm setting, \correctionmodule iteratively prompts the model to repair the sample workflow for its defect incidences per category till all reported incidences for the target defect type are repaired or the \maximumcountofrepairattempts is exhausted. \rev{This means that \correctionmodule repairs defect incidences across varying types sequentially. If \correctionmodule is able to repair defect incidences of a certain type before or by the time the maximum count of repair attempts is reached, the repair of the target defect type is considered to be successful. Based on prior studies conducting iterative examination to account for the stochasticity in \fm completions~\cite{CodeRL2022Le, lyu2025top},} we set the \maximumcountofrepairattempts for each defect type to \attemptcount and compare the performances of the used models with each other in
% for 
repairing incidences of each defect type as a value of \passk accuracy~\citep{chen2021evaluating}. In this context, for each defect type, the model with the best repair capabilities is the one with the highest number of correctly repaired defect incidences from the initial count of reported defects by \staticanalyzer given at most ten (i.e., k=$10$) repair attempts for that particular defect type.

After identifying the best performing \fm,
% with the optimum defect repair performance,
we implement the test-time strategy for that specific model. Specifically, \correctionmodule iteratively prompts the selected \fm to repair defect incidences of a particular type, while calculating the repair performance of the \fm at each iteration using pass@k accuracy. To evaluate the effect
% influence 
of increasing attempts on the \fm{}'s repair capabilities we increase k (ranging from one to ten) and investigate whether increasing the number of repair attempts 
% counts would 
enhances the \fm{}'s repair capabilities.
% \rev{For each \fm,} we \rev{implement the test-time strategy by iteratively prompting the model to} repair the sample \rev{workflow} for \rev{its} defect incidences per category till all reported incidences for the target defect type are repaired or the \maximumcountofrepairattempts are exhausted. 
% To account for nondeterministic behavior of \fms, 
% We set the \maximumcountofrepairattempts for each defect incident to \attemptcount and report the performance of \correctionmodule as \passk accuracy \rev{with \textit{k} ranging from one to ten}~\citep{chen2021evaluating}, that is, the number of correctly repaired defect incidences from the initial count of reported incidences for the target category by \staticanalyzer given k repair attempts for each incident.

To minimize the confounding factors and to be able to evaluate the effectiveness of \correctionmodule in correcting reported defect incidences, \correctionmodule is executed for each defect category independently from the rest of the defect types. This means, for instance, after the sample workflow is attempted for the repair of all 
% of 
its \undefvar, 
% defect incidences,
to repair the reported incidences of \unuse, we use the initial defective workflow sample as \correctionmodule{}'s input 
% the 
% baseline 
% input \res{of}
%  for 
% \correctionmodule 
and not the version of the workflow that has been repaired for its unreachable variable
defects.
% \undefvar 

\rev{\correctionmodule can introduce new defects while repairing defective \workflow{}s. The newly introduced defects by \correctionmodule, however, will be detected by \staticanalyzer as after each repair attempt by \correctionmodule, the target workflow is re-examined by the static analyzer to verify whether all original defect incidences are resolved. %, i.e.,  calculating \correctionmodule{}’s repair performance . 
As such, if \correctionmodule generates new defects while trying to repair the original incidents, the newly occurring defects will be detected and localized by \staticanalyzer, with their detection report sent back to \correctionmodule for initiating their repair. Conceptually, \correctionmodule can also generate defects that are not detectable by \staticanalyzer, i.e., defect types such as omission of logic or violation of concurrency constraints that require human intervention or the execution of workflows for their detection, respectively. However, we have yet to observe such errors caused by \correctionmodule in practice. %, when the tool is used for repairing \workflow{}s.
}

Prior research demonstrated that \fms are incapable of conducting intrinsic self-correction attempts, solely based on their inherent capabilities and no use of external feedback, for improving their initial generation accuracy~\cite{kim2023language, huang2023large}. Interestingly, the performances of these models are even shown to degrade after a few iterations of self-correction attempts. Therefore, we design \correctionmodule to incorporate direct feedback from \staticanalyzer{}'s defect reports for an accurate defect repair performance. Nonetheless, we conduct an evaluation on the effectiveness of incorporated feedback from \staticanalyzer on the repair performance of \correctionmodule. To do so, we directly prompt \correctionmodule to analyze an input \workflow for existing defect incidences and repair them, if any, in place. The definition of 
% plausible 
defect types and corresponding few-shot examples are appended as additional context information to the prompt instructions. Similar to the incorporation of feedback, this setting is also implemented as a test-time iterative strategy, with pass@k (k=$10$) calculated as the accuracy of \correctionmodule.

% In the following two sections, we described the mechanisms of \staticanalyzer and \correctionmodule and their functionalities in detail.

\subsection{Experiment Results}
\label{sec: results}
In this section of the paper, we present the results of the experiment. We set out to find answers to the following two research questions.

\par\smallskip\noindent{\textbf{\reqtwo}}

We develop \staticanalyzer as an automated detection tool assisting developers to identify defect incidences in \fm{}-\rev{generated} %synthesized 
\dsl programs. Hence, we assess how accurately \staticanalyzer can locate defects within a \workflow{}.

\par\smallskip\noindent{\textbf{\reqthree}}

We conclude our effort in automatic detection of defects in \workflow{}s by attempting to repair them using our repair tool, \correctionmodule. Hence, we also assess the effectiveness of \correctionmodule in repairing the detected defects by \staticanalyzer.

\bighead{Accuracy of Detection~(\justrqtwo)}
\label{sec:detection-accuracy}
The accuracy of \staticanalyzer in detecting defect incidences in %\rev{both synthesized and real-world \openagi} 
\workflow{}s is demonstrated in \autoref{tab:static-analyzer-performance}. The accuracy of the tool varies across defect categories, with 100\% precision and recall for the detection of \badjson, \baddsl, \hallucinatedskills, \badparameter, \badexp, and \badoutput \rev{across the \samplesize simulated scenario-based workflows.} %synthesized from simulated user scenarios}. 
\rev{Similarly, across the \openagi workflows, \staticanalyzer is able to detect \baddsl, \hallucinatedskills, \badparameter, and \badoutput with 100\% precision and recall.} \rev{It should be mentioned that, as no defect incidence of \badjson and \badexp exist across either of simulated scenario-based or \openagi workflows, for the sake of conducting a complete evaluation of the detection capabilities of \staticanalyzer, we randomly choose ten simulated scenario-based workflows to change their structure such that they do not parse into \json objects anymore. As such, we can investigate \staticanalyzer{}'s performance in detecting \badjson{}s. Similarly, we randomly select ten additional simulated scenario-based workflows and inject them with \badexp (one incident per workflow) to evaluate \staticanalyzer{}'s detection performance for this defect type. We do not choose any of the \openagi \workflow{}s to inject them with incidences of \badjson or \badexp, as these workflows are generated from real-world scenarios. Artificially manipulating \openagi workflow structures compromises their representative nature of %\openagi workflows %of the second evaluation fold.} 
% of 
real-world settings.} 

\staticanalyzer demonstrates an absolute recall and a high precision (75\%) when attempting to detect \unuse \rev{across simulated scenario-based workflows}. %of the first evaluation fold}. 
\rev{Interestingly,} \staticanalyzer has almost identical recall when detecting \undefvar and \badtype (71\% and 72\%, respectively) \rev{across the same workflows}. However, its precision is different 
with $50$\% and $37$\% for \undefvar and \badtype, respectively. 
\rev{\staticanalyzer demonstrates similar detection performance for unreachable and unused variables across simulated scenario-based and \openagi workflows.} 
% \rev{\staticanalyzer{}'s performance in detecting unreachable and unused variables across \openagi \workflow{}s is on par with its detection performance for the same defect types across synthesized \workflow{}s.}
For detecting %both of 
% these \rev{three} defect types 
\rev{unreachable and unused variables and \badtype}, all program paths in the input \workflow{}s should be analyzed to detect variable placements on infeasible paths. Analyzing the program paths for their feasibility requires evaluating all possible combinations for input parameters,
% combinations, 
even when their values are only known at runtime. Hence, additional measures alongside static analysis should be placed to enable such an evaluation of program paths. This necessity is discussed in more detail in \autoref{sec:pi-1}. 
% In order to further analyze the differences between \staticanalyzer{}'s performance in detecting \undefvar and \badtype, we apply \fisherexact~\citep{fisher1922interpretation} to investigate the potential significant difference between the population of \tp{}s and \fp{}s across these two defect types. 
% The conducted test is non-parametric, similar to many other prior conducted studies~\citep{moussa2022meg, maipradit2023repeated}, as we do not make any assumptions regarding the probability distribution of the data under investigation. Furthermore, as we are conducting one comparison, no correction on the threshold of significance is necessary~\citep{armstrong2014Bonferroni}. Hence, the significance threshold is set to $0.05$. Based on the result of the test (i.e., $p-value = 0.21$) we cannot infer any statistically meaningful differences between the ability of \staticanalyzer in detecting \undefvar and \badtype.

\begin{myframe}[width=\linewidth, top=0pt,bottom=0pt,left=0pt,right=0pt,arc=0pt,auto outer arc]
\small{\textbf{\underline{\textit{\justrqtwo{}.}}} Although having varied performance across different categories, \staticanalyzer demonstrates an acceptable performance in detecting defect incidences of \workflow{}s, with a comparatively high recall of $80\%$ \rev{and $87\%$ across simulated scenario-based and \openagi workflows, respectively.} %of simulated scenarios and workflows of the real-world \openagi scenarios, respectively}. 
This is particularly important, as \staticanalyzer is designed to detect the majority of existing defects with high confidence.}
% accuracy varies across defect types with its precision and recall ranging from $37\%-100\%$ and $71\%-100\%$, respectively.}
\end{myframe}

% \usepackage{booktabs}
% \usepackage{graphicx}
\begin{table}[!htp]
\centering
\footnotesize
\setlength{\tabcolsep}{1pt}
\captionsetup{justification=centering}
\caption{\staticanalyzer accuracy across different defect categories}
\label{tab:static-analyzer-performance}
% \resizebox{0.6\columnwidth}{!}{ %

\begin{tabular}{@{}lrrrr@{}}
\toprule
          \textbf{Workflow Type}                                   & \multicolumn{2}{c}{\textbf{Shadow}}                                               & \multicolumn{2}{c}{\textbf{\openagi}}                                               \\ \midrule
\textbf{Defect Category}                     & \multicolumn{1}{c}{\textit{\textbf{Precision}}} & \multicolumn{1}{c}{\textit{\textbf{Recall}}} & \multicolumn{1}{c}{\textit{\textbf{Precision}}} & \multicolumn{1}{c}{\textit{\textbf{Recall}}} \\ \midrule
\textbf{\cbadjson}            & 100\%                                           & 100\%                                        & N/A                                             & N/A                                          \\
\textbf{\cbaddsl}             & 100\%                                           & 100\%                                        & 100\%                                           & 100\%                                        \\
\textbf{\cundefvar}           & 50\%                                            & 71\%                                         & 33\%                                            & 58\%                                         \\
\textbf{\cunuse}              & 75\%                                            & 100\%                                        & 53\%                                            & 89\%                                         \\
\textbf{\cbadtype}            & 37\%                                            & 72\%                                         & N/A                                             & N/A                                          \\
\textbf{\challucinatedskills} & 100\%                                           & 100\%                                        & 100\%                                           & 100\%                                        \\
\textbf{\cbadparameter}       & 100\%                                           & 100\%                                        & 100\%                                           & 100\%                                        \\
\textbf{\cbadexp}             & 100\%                                           & 100\%                                        & N/A                                             & N/A                                          \\
\textbf{\cbadoutput}          & 100\%                                           & 100\%                                        & 100\%                                           & 100\%                                        \\ \midrule
\textbf{Overall}                             & 56\%                                            & 80\%                                         & 67\%                                            & 87\%                                         \\ \bottomrule
\end{tabular}
% }%
\end{table}

% \noindent\paragraph{\underline{\textbf{Effectiveness of repair (\justrqthree)}}}
\bighead{Effectiveness of Repair~(\justrqthree)}
\label{sec:repair-performance}
\autoref{tab:repair-module-performance} provides granular pass@10 metrics for the count of repaired defect incidences across different splits of executing \correctionmodule \rev{over the \samplesize simulated scenario-based workflows,} using \gpt{}-4o, \llama{}, and \qwen models. As can be observed from the table, the best collective performance belongs to \gpt{}-4o, with \llama and \qwen performing $75.17$\% and $85.66$\% worse, respectively.

% In the preamble
% \usepackage{booktabs}
% \usepackage{graphicx}
% \usepackage{makecell}

\begin{table}[!ht]
\centering
\captionsetup{justification=centering}
\caption{Defect repair performance of \correctionmodule using three different FMs (GPT-4o, Llama, Qwen). Results demonstrate the improvement gained from static analysis feedback.
}
\label{tab:repair-module-performance}
\resizebox{\linewidth}{!}{%
\begin{tabular}{@{}lrrrrrrrrr@{}}
\toprule
\multicolumn{1}{c}{\textbf{}}             
& \multicolumn{1}{c}{\makecell{\textbf{Detection}\\\textbf{Count}}}
& \multicolumn{4}{c}{\makecell{\textbf{Pass@10}\\\textbf{Repair Count}}}
& \multicolumn{4}{c}{\makecell{\textbf{Relative}\\\textbf{Improvement (\%)}}} \\ 
\midrule
\textbf{Defect Categories}                
& \multicolumn{1}{c}{\textit{\textbf{Timon}}}  
& \multicolumn{1}{c}{\textit{\textbf{Baseline}}} 
& \multicolumn{1}{c}{\textit{\textbf{\gpt{}-4o}}} 
& \multicolumn{1}{c}{\textit{\textbf{\llama}}} 
& \multicolumn{1}{c}{\textit{\textbf{\qwen}}} 
& \multicolumn{1}{c}{\textit{\textbf{Baseline}}} 
& \multicolumn{1}{c}{\textit{\textbf{\gpt{}-4o}}} 
& \multicolumn{1}{c}{\textit{\textbf{\llama}}} 
& \multicolumn{1}{c}{\textit{\textbf{\qwen}}} \\ 
\midrule
\textbf{Invalid DSL}                      & 3   & 0  & 0   & 0   & 0   & 0\%   & 0\%   & 0\%   & 0\%   \\
\textbf{Unreachable variables}            & 254 & 7  & 20  & 21  & 7   & 2.76\%& 7.87\%& 8.27\%& 2.76\%\\
\textbf{Unused variables}                 & 128 & 4  & 60  & 16  & 0   & 3.13\%& 46.88\%& 12.50\%& 0\%   \\
\textbf{Incorrect data type propagation}  & 55  & 0  & 12  & 0   & 0   & 0\%   & 21.82\%& 0\%   & 0\%   \\
\textbf{Hallucinated skills}              & 3   & 3  & 3   & 3   & 3   & 100\% & 100\% & 100\% & 100\% \\
\textbf{Skills with defective parameters} & 516 & 0  & 160 & 0   & 0   & 0\%   & 31.01\%& 0\%   & 0\%   \\
\textbf{Incorrect outputs}                & 31  & 31 & 31  & 31  & 31  & 100\% & 100\% & 100\% & 100\% \\
\midrule
\textbf{All defect types}                 & 990 & 45 & 286 & 71  & 41  & 4.55\%& 28.89\%& 7.17\%& 4.14\%\\
\bottomrule
\end{tabular}%
}
\end{table}

\llama cannot repair any incidences of \badtype and \badparameter while the accuracy of \gpt{}-4o for repairing these defects is $21.82$\% and $31.01$\%, respectively. Although repairing some incidences of \unuse, \llama{}'s performance is $34.37$\%
% $12.50$\% 
points lower than \gpt{}-4o's. The performance of \llama is on par with \gpt{}-4o's, when repairing \badoutput, \hallucinatedskills, \baddsl, and \undefvar. 

Compared to \llama, \qwen performs even worse. \qwen cannot repair any of \badtype, \unuse, \badparameter, and \baddsl. When repairing \undefvar, \qwen{}'s performance is $5.12$\% points lower compared to that of \gpt{}-4o, only repairing seven out of $254$ detected incidences. Finally, \qwen{}'s performance is on par with \gpt{}-4o's, when repairing \badoutput and \hallucinatedskills.

\rev{\autoref{tab:repair-module-performance} shows that regardless of the used \fm, \correctionmodule is incapable of repairing \baddsl{}s. The reason for this can be attributed to the limited capabilities of \fm{}s in generating structured outputs~\cite{shorten2024structuredrag}. Although literature focuses on evaluating the abilities of \fm{}s in directly synthesizing structured outputs, repairing a defective \dsl can also be considered as re-synthesizing the program such that it correctly adheres to the structure of the specified grammar. This, in turn, explains why \correctionmodule does not repair \baddsl{}s successfully. On the other hand, we observe that in all settings of the underlying \fm, \correctionmodule repairs the incidences of \hallucinatedskills. Compared to the repair of complex defects such as \baddsl{}s, the repair of \hallucinatedskills is much simpler. The incidences of \hallucinatedskills pertain to already existing actions in the skill database, that are referenced with faulty names (i.e., spelling errors or incorrect format) in the workflows. In other words, no hallucination occasion exists in the end where the synthesizer \fm had fabricated an action in the generated workflow in the beginning. As the existing skills are appended to the context of the repair prompt (Section~\ref{sec: correction-module}), \correctionmodule{}'s underlying \fm can compare the actions in the target workflow with their actual forms, correcting the names of the hallucinated skills.}

From these results, we observe that model performance varies across models, which may be attributed to factors such as model architecture (MoE vs. Dense), parameter count, and training data composition. For instance, the results demonstrate that the higher the count of the learnable parameters in an LLM, the better its performance in repairing the defect incidences in \workflow{}s.

\autoref{tab:repair-module-performance} also demonstrates the effectiveness of incorporated feedback from \staticanalyzer{}'s defect incidences report on the repair performance of \correctionmodule. Specifically, the capabilities of \gpt{}-4o in directly detecting and repairing defect incidences in \workflow{}s are shown as the baseline repair accuracy at pass@10. In this setting, \correctionmodule is able to repair only $4.55$\% of all defect incidences across the evaluation set which is $84.25$\% relatively lower repair performance compared to when \staticanalyzer{}'s feedback is also incorporated. According to these results and all prior comparisons between different \fm{}s' capabilities in repairing defect incidences, \gpt{}-4o with incorporated feedback from \staticanalyzer is demonstrated to be \correctionmodule{}'s best performing setting in repairing defective \workflow{}s. Hence, the test-time iterative strategy is implemented for this setting. \rev{Additionally, the same configuration (i.e., \gpt{}-4o incorporated with \staticanalyzer{}'s feedback) is used for investigating \correctionmodule{}'s repair performance across the \openagi workflows.} %\workflow{}s.} 
% that are representative of real-world settings (i.e. \openagi workflows).}

\begin{figure}[t]
    \centering
    \includegraphics[width=0.99\linewidth]{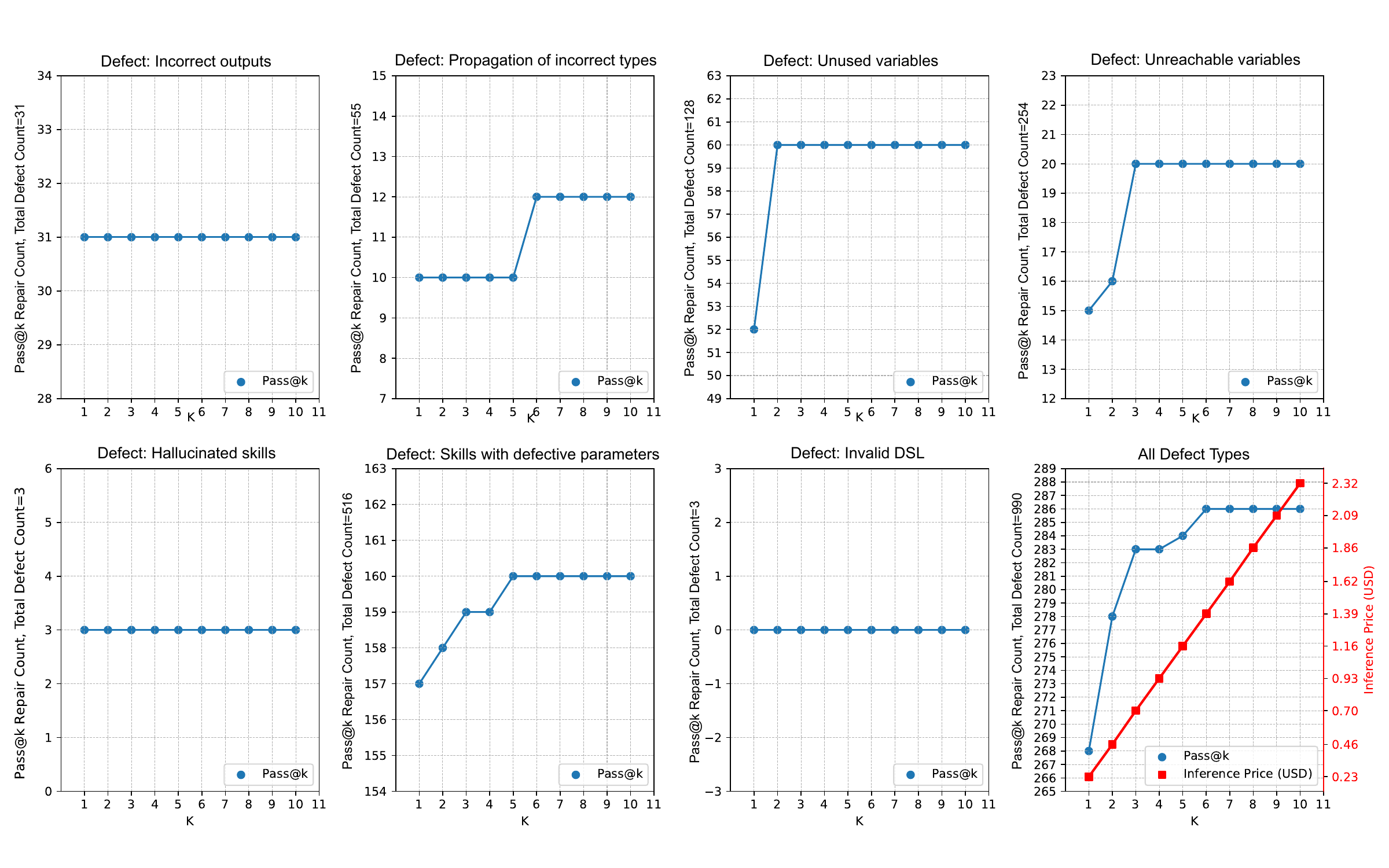}
    \caption{Effect of test-time iterative strategy on \correctionmodule}
    \label{fig:test-time-pumbaa}
\end{figure}

\autoref{fig:test-time-pumbaa} illustrates the results of the test-time iterative strategy for \gpt{}-4o as pass@k accuracies, across different values of k ranging from one to ten. As observed from the figure, increasing k improves the repair performance of \correctionmodule. From the six repaired defect types, \correctionmodule repairs all of the \badoutput and \hallucinatedskills at pass@1. For the remaining four defect types, \correctionmodule{}'s performance saturates at pass@2, pass@3, pass@5, and pass@6 for \unuse, \undefvar, \badparameter, and \badtype, respectively. Although \correctionmodule{}'s performance does not saturate uniformly across different defect types, the collective trend illustrates a direct relationship between iteration and performance. Specifically, $27.07$\%, $28.08$\%, $28.59$\%, $28.59$\%, and $28.69$\% of all defect incidences are repaired at pass@1, pass@2, pass@3, pass@4, and pass@5. Finally, the performance saturates by the $6$\textsuperscript{th}
% sixth 
pass, repairing $28.89$\% of all defect incidences which account for $10$\% of the \rev{population of simulated scenario-based workflows} %evaluation \rev{fold} %set 
as fully repaired \rev{programs}. %\workflow{}s. 

\rev{\autoref{fig:test-time-pumbaa} also demonstrates the corresponding costs for repairing the incidences of all defect types across the simulated scenario-based workflows and varying repair attempts. The cost of each repair attempt is calculated as the sum of the monetary cost corresponding to input token usage and completion token count for that specific repair attempt. The inference cost of \gpt{}-4o is USD 2.5 and USD 10 per one million input and completion tokens, respectively~\cite{openai_pricing}.
As expected, increasing the number of repair attempts directly increases costs (i.e., a linear relationship). Given the saturation of repair performance at pass@6 and the ever-increasing token usage and corresponding costs, we recommend setting \correctionmodule{}'s maximum repair attempt count to six in practical settings. However, based on their discretion, users may choose a different value for the maximum allowed repair attempts.}

\rev{\autoref{tab:defect-repair-openagai} demonstrates \correctionmodule{}'s performance in repairing defect incidences of the \openagi \workflow{}s. It is evident that the performance of \correctionmodule in repairing \hallucinatedskills and \badoutput is the same as the tool’s performance when repairing the same defect types across the simulated scenario-based workflows. %\workflow{}s. 
Additionally, \correctionmodule{}’s repair performance for \undefvar in \openagi workflows %\workflow{}s 
is on par with the tool’s performance when it is employed to repair the same defects %\undefvar 
in simulated scenario-based workflows. %\workflow{}s 
% with simulated user scenarios. 
Compared to \correctionmodule{}’s performance in repairing incidences of \badparameter and \baddsl{}s in simulated scenario-based workflows, the tool’s performance demonstrates considerable enhancement when repairing the same defect types across the \openagi workflows. Particularly, the tool’s performance increases from 31.01\% to 100\% and from 0\% to 12.50\% while repairing incidences of \badparameter and \baddsl{}s, respectively. On the other hand, \correctionmodule{}’s performance in repairing \unuse decreases 20.21\% points, when the tool is applied to the \openagi workflows %\workflow{}s 
instead of the simulated scenario-based ones. %workflows. 
Generally, however, the repair performance of \correctionmodule increases 11.59\% points when it is applied to the \openagi workflows %\workflow{}s 
instead of the simulated scenario-based workflows, demonstrating its applicability in practical settings.}

\begin{table}[htbp]
\centering
\footnotesize
\captionsetup{justification=centering}
\setlength{\tabcolsep}{1pt}
\caption{Defect repair performance of \correctionmodule across the \openagi \workflow{}s}
\label{tab:defect-repair-openagai}
% \resizebox{0.70\columnwidth}{!}{ %
\begin{tabular}{lr@{\hspace{0.5cm}}r@{\hspace{0.5cm}}r}
\toprule
\textbf{Defect Categories} & \textbf{\begin{tabular}[c]{@{}c@{}}Detection Count\\ (\staticanalyzer)\end{tabular}} & \textbf{\begin{tabular}[c]{@{}c@{}}Pass@10\\ Repair Count\end{tabular}} & \textbf{\begin{tabular}[c]{@{}c@{}}Relative\\ Improvement (\%)\end{tabular}} \\
\midrule
\textbf{\cbaddsl} & 8 & 1 & 12.50\% \\
\textbf{\cundefvar} & 33 & 3 & 9.09\% \\
\textbf{\cunuse} & 15 & 4 & 26.67\% \\
\textbf{\challucinatedskills} & 15 & 15 & 100\% \\
\textbf{\cbadparameter} & 12 & 12 & 100\% \\
\textbf{\cbadoutput} & 4 & 4 & 100\% \\
\midrule
\textbf{All defect types} & 84 & 34 & 40.48\% \\
\bottomrule
\end{tabular}
% }%
\end{table}

\begin{figure}[t]
    \centering
    \begin{subfigure}{0.5\columnwidth}
        \centering
        \includegraphics[
            width=\linewidth,
            height=0.25\textheight,
            keepaspectratio
        ]{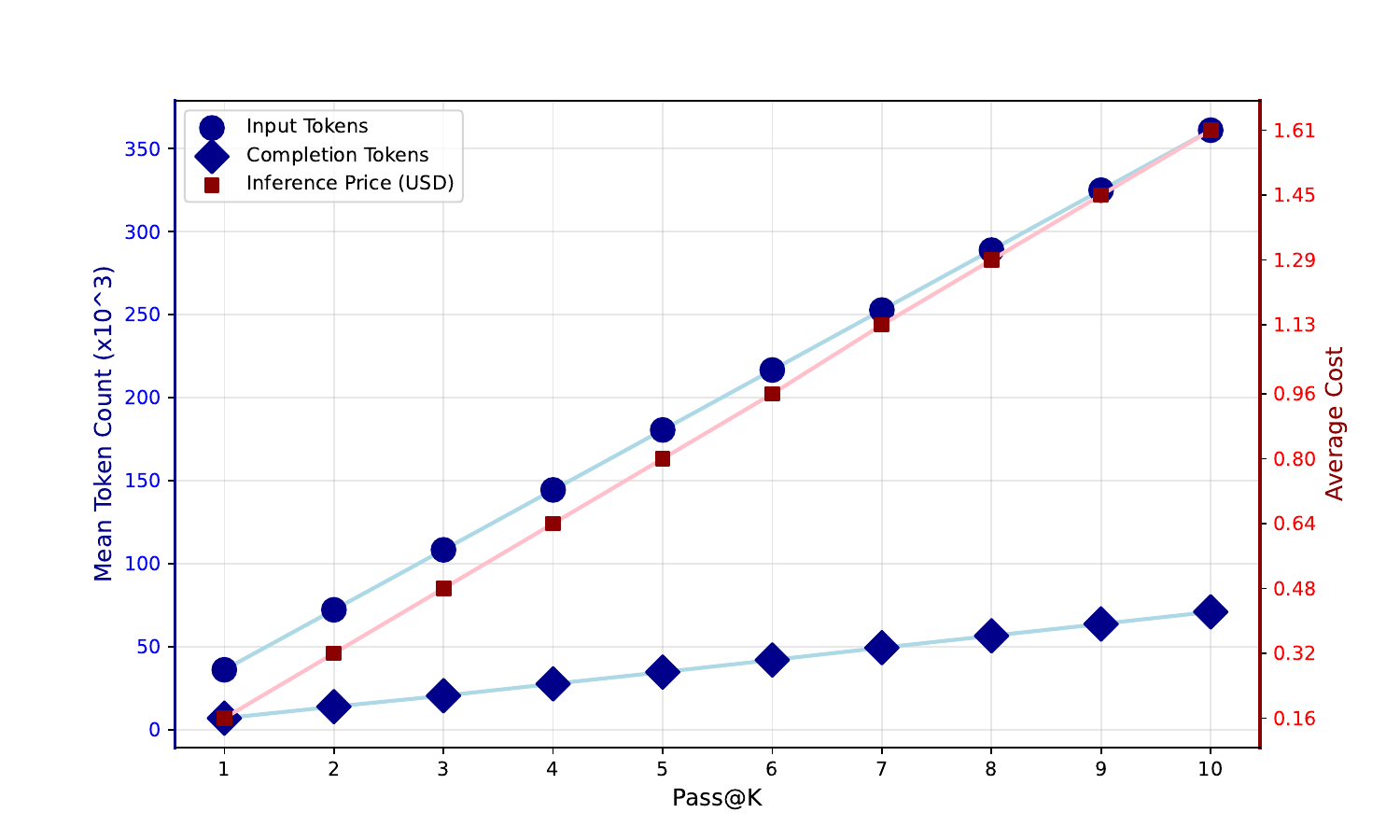}
        \caption{Token and monetary cost}
    \end{subfigure}%
    \begin{subfigure}{0.5\columnwidth}
        \centering
        \includegraphics[
            width=\linewidth,
            height=0.25\textheight,
            keepaspectratio
        ]{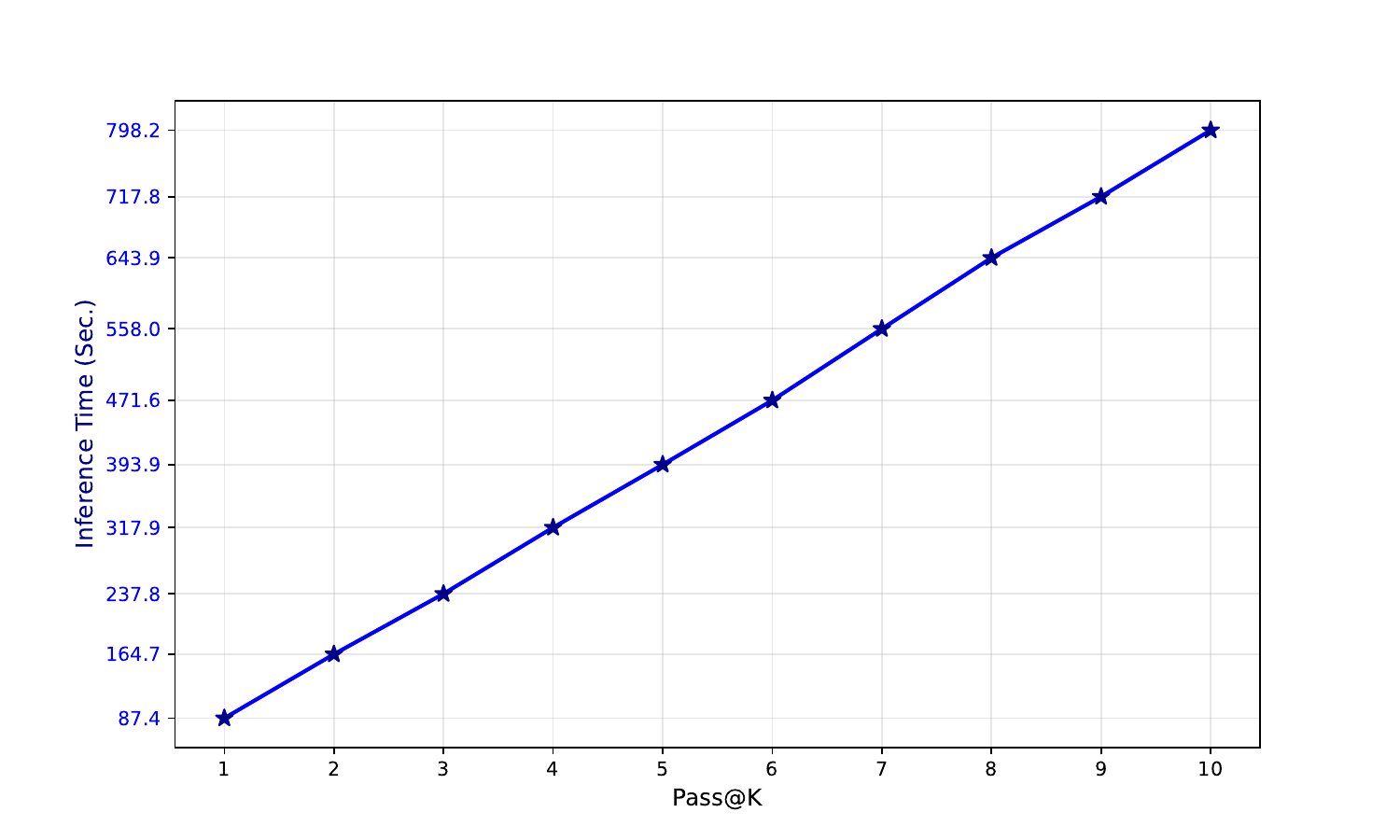}
        \caption{Inference time}
    \end{subfigure}

    \caption{Token, cost, and time analyses for the repair of the \openagi \workflow{}s}
    \label{fig:monetary-cost-openagi}
\end{figure}

\rev{We also conduct token usage, monetary cost, and inference time analyses for the repair of defective \openagi workflows by \correctionmodule. As can be observed from \autoref{fig:monetary-cost-openagi}, similar to the case of applying \correctionmodule to repair defect incidences across the simulated scenario-based workflows, %\workflow{}s, 
with each repair attempt, the token usage and the corresponding cost increase linearly.}

% \begin{figure}[t]
%     \centering
%     \includegraphics[width=0.7\linewidth]{figures/wfs_workflow_59--workflow_68_new_major_cost_analysis.pdf}
%     \caption{Token, cost, and time analyses for the repair of the \openagi \workflow{}s}
%     \label{fig:monetary-cost-openagi}
% \end{figure}

Although not all existing defect incidences in the evaluation set are repaired, using \correctionmodule in a production setting can still be valuable. \correctionmodule{}'s accuracy in repairing \badoutput and \hallucinatedskills is $100$\% \rev{across both simulated scenario-based and real-world \openagi workflows}. This is especially important as, in practice, not all \workflow{}s are tainted with other defect types, for which \correctionmodule{}'s repair performance may be lower, making \correctionmodule a practical tool for 
% industrial 
production settings. Additionally, \correctionmodule is only one part of our end-to-end pipeline for generating robust and executable
% automatic 
workflows using \fms. If \correctionmodule is not able to repair all defect incidences in input \workflow{}s, to be feasibly used in production, the system should incorporate fallback mechanisms. For instance, after showing a detailed compilation of remaining defect types and their locations in the input \workflow{}s, the system should defer to end users and their expertise for repairing them.
% the remaining defect incidences.
% after demonstrating a detailed compilation of still existing defect types and their locations in the input \workflow{}s.}

% with the repair of \badoutput and \hallucinatedskills defects for which all incidences are fully resolved---100\% relative improvement compared to the detection baseline. On the other hand, \correctionmodule is unable to correctly restructure any of the invalid \dsl program instances---no improvement compared to baseline. Similarly, \correctionmodule demonstrates varying abilities in repairing unreachable and unused variables and \badtype{}s improvement ranging between $7.87$\% and $46.88$\%. 
% As such, we can conclude that the ability of \fms in repairing defective \workflow{}s is confounded by the types of their defect incidences.    

\begin{myframe}[width=\linewidth, top=0pt,bottom=0pt,left=0pt,right=0pt,arc=0pt,auto outer arc]
\small{\textbf{\underline{\textit{\justrqthree{}.}}} The ability of \fms to repair the defects of \fm{}-\rev{generated} %synthesized 
\dsl programs varies across different families of models and different defect types. Nonetheless, incorporating feedback from \staticanalyzer{}'s defect incidence report and implementing a test-time iterative strategy have been shown to enhance \correctionmodule{}'s repair performance.}
% Specifically, \fms seems capable of reformatting outputs field based on specific instructions. On the other hand, they demonstrate difficulties in restructuring the workflows in accordance with the specified \dsl grammar.}

\end{myframe}

\section{Discussion and Practical Implications}
\label{sec:discussion}
In this section of the paper, we discuss the results and their implications for researchers and developers.

\subsection*{\practicalimplication{} \textit{1}: Practitioners should consider additional mechanisms to successfully synthesize \dsl program instances using \fms.}
Our observations demonstrate that even the most powerful proprietary \fms, with MoE architecture and a larger count of learnable parameters, are not successful in directly synthesizing \dsl workflows. Hence, employing additional guardrails for ensuring the \fms{}' success in synthesizing \dsl program instances is inevitable. Specifying the \dsl as a \fm{}-friendly grammar should be the initial consideration to be explored for enabling a successful synthesis of workflows by \fms. Additionally, the results of our implemented test-time iterative strategy demonstrate the direct relationship between the repair iteration attempts and the performance of \fms in repairing the defect incidences in \workflow{}s. With more attempts, the \fms repair more defect incidences until their performance is saturated. Similarly, it can be expected that implementing the test-time iterative strategy can also result in \fms{}' higher success rate in synthesizing \dsl program instances. Finally, based on the results of our study, even increasing the synthesis attempt may not be 
% solely 
enough for the state-of-the-art \fms to generate a correct \dsl program instance without any defect incidences. This emphasizes the need for implementing extra mechanisms
% (other than only focusing on \fms for maximizing their synthesis success rate)
% a focused effort on the \fms for maximizing their capabilities for a higher synthesis success rate---
such as static and dynamic analyses to detect any defect incidences that may have found their way into the synthesized \dsl program instance, despite capitalizing on \fms to achieve optimal synthesis accuracy.

\subsection*{\practicalimplication{} \textit{2}: Static analyzers require additional support to effectively detect all defects in FM-\rev{generated} \dsl programs.}\label{sec:pi-1}
Our defect detection tool \staticanalyzer{}, detects incidences of \detectcount defect categories in \workflow{}s by operating two components: \textit{Structure Validator} and \textit{Defect Detector} (modules~\bluecircled{A} and~\greencircled{C} in Section~\ref{sec:static-analyzer}, respectively). The former component examines whether a \workflow is structured as a valid \json and whether it adheres to the specified \dsl grammar. 
% Ensuring the existence of utilized skills by the \fm for implementing \taskaction{}s and also the correctness of their input and output parameters is conducted through the comparison of the target skill with the hash-map of all available skills which is also considered as a light weight analysis.
This component is able to detect all incidences of the relevant defects with $100\%$ accuracy. The latter component uses reconstructed
% reconstructs 
% for detecting unreachable and unused variables and \badtype{}s, the \staticanalyzer 
% the 
complex CFG
% control flow constructs 
of \workflow{}s and the subsequent
% to build 
\usedefdefuse to detect the defects in workflows. In contrast to the first component's perfect performance, implemented static analyses in the current
% second 
component are not able to detect all defect incidences. Further investigation of unsuccessful detection instances reveals that similar to programs written in general-purpose programming languages, \fm{}-\rev{generated} %synthesized 
\dsl programs are also prone to \textit{infeasible program execution paths} when the program involves non-linear control flow constructs and variables whose values are only known at runtime.
% complex data and control flow errors. Specifically, We identify \textit{mutually exclusive conditions}
% or impossible program paths 
% across our dataset. 
These incidences pertain to \textit{possibly unreachable} and \textit{possibly unused} variables or variables with definitions and use cases located in occasionally inaccessible program paths. 

\begin{figure}[t]
\centering
\footnotesize
% \begin{minted}[highlightlines={3},highlightcolor=yellow]{ruby}
\begin{lstlisting} [language=wf, escapechar=!, caption={A possibly unreachable variable in a simplified FM-generated DSL workflow instance}, label=possibly-unreachable]
workflow: {
tasks: [
  { id: "Multiply_or_Sum_Two_Values",
    if: {
      condition: "x > 10 && y < 5",
      then: ["p=xy"],
      else: ["q=x+y"]}},
  { id: "Multiply_or_Divide_by_2",
    if: {
      condition: "x < 15 && y > 0",
      then: ["s=p/2"],
      else: ["t=2q"]}}]}

\end{lstlisting}  
\end{figure}

Consider \autoref{possibly-unreachable} as an example.
In this example, investigating the
% variable 
runtime bounds of variables and their relationships is necessary to detect mutually exclusive and contradictory conditions and to verify the satisfiability of all 
% their 
possible combinations for variable values. Loops are also affected by runtime bounds. Only at runtime can we determine whether a loop has at least one iteration or its termination condition is satisfied. 
\rev{The same applies to \badtype. For instance, if variable ``v'' is defined as a float and then used as an integer in a subsequent node, \staticanalyzer only detects the \badtype with full confidence if the variable use occurs on a path that is always executed. Otherwise, Timon may miss the defect.}
Therefore, sole static analysis cannot confidently detect whether any defect exists. For detecting such defect occurrences in general-purpose programs, dynamic analysis or \csmt~\citep{nguyen2013semfix, srivastava2010program, de2008z3} are applied to investigate whether the solution space can satisfy all possible combinations of runtime bounds and conditions. Similarly, to accommodate the detection of such defects in FM-\rev{generated} %synthesized 
\dsl programs, we suggest incorporating \smt{} solvers alongside static analyzers.

\subsection*{\practicalimplication{} \textit{3}: Prompting assistance is needed to improve the capabilities
% abilities 
of \fms in \dsl synthesis.}
We develop \correctionmodule to repair defect incidences in \fm{}-\rev{generated} %synthesized 
\dsl programs with the assistance of FMs. For this purpose, we provide \correctionmodule{}'s \fm with \staticanalyzer{}'s defect reports as direct feedback to repair defect incidences correctly and comprehensively.
% hints for guiding the \fm{}'s generation towards a complete repair of defects. 
% To ensure of the quality of the generation, 
As \correctionmodule{}'s \fm, we settle on \gpt{}-4o, a \fm with complex abilities, suitable for solving intricate problems such as program synthesis, verification, and repair~\citep{wen2024enchanting}. Furthermore, to ensure our choice for \correctionmodule{}'s \fm, we conduct experiments comparing the performance of \gpt{}-4o with other \fms, demonstrating the superior performance of \gpt{}-4o in repairing defect incidences compared to the rest of the investigated \fms. Additionally, we ensure to improve \correctionmodule{}'s performance by iteratively refining its prompts starting from carefully handcrafted versions, rewriting for clarity, adding few-shot examples, and refactoring them based on \openai{}'s and \an{}'s guidelines, verifying that the performance is increased on a subset of defective samples with every enhancement. 
% prompt the \fm with detailed instructions, accompanying various few-shot examples~\citep{brown2020language}.
% for clearly describing all possible situations including edge-cases . 
% We handcraft our prompts through multiple iterations to ensure their coherence, correctness, clarity, and comprehensiveness. 
% We continue refining these prompts, till an engineer external to the team and unbiased to the problem at hand, verifies the quality of the prompts.
% In spite of our best efforts, the results of the experiment conducted for evaluating the effectiveness of \fms 
Nevertheless, \correctionmodule falls short of our initial expectations in repairing \fm{}-\rev{generated} %synthesized 
\dsl programs (see Section~\ref{sec:repair-performance}).
% is not as good as our initial expectations. 
We believe that prompts are one potential area for improvement to address this marginal performance.
Despite their proven effectiveness in similar problems~\citep{austin2021program, ahmed2022few}, no prior research has explored how effective the current state-of-the-practice prompting techniques are in guiding \fms in repairing FM-\rev{generated} %synthesized 
\dsl programs. Hence, we conduct a \textit{pilot exploratory investigation} on using automated \textit{prompt optimization tools}, specifically \anthropic~\citep{anthropic_prompt_improver}, to improve \correctionmodule{}'s performance. Recent advancements in these optimization tools demonstrate promising trajectories for improving the performance of \fms in tackling complex tasks. Hence, we hypothesize that using these optimizers can also improve the performance of the \fms in repairing FM-\rev{generated} %synthesized 
\dsl programs. However, the results of our analysis of the effectiveness of the automated prompt optimizer are mixed, ranging from a 38.33\% decrease in repair capability for \badparameter, to no change for \badoutput, \badtype, and \hallucinatedskills, to a 55\% improvement in repairing \undefvar and \baddsl. Consequently, we suggest further exploration of the benefits of incorporating different prompt optimizers, such as \dspy~\cite{khattab2024dspy} and \textsf{TextGrad}~\cite{yuksekgonul2024textgrad}, for refining the \fms{}' prompts to increase their capabilities of repairing defective \fm{}-\rev{generated} %synthesized 
\dsl programs.
% plan to explore the application of different prompt optimizers such as \dspy~\cite{khattab2024dspy} and TextGrad~\cite{yuksekgonul2024textgrad}, for refining the effectiveness of the prompts, further in future.
% A potential reason for this marginal-effectiveness of \fms in repairing these programs can be related to our utilized prompting techniques. Using \fms for \dsl program verification and repair is an undiscovered territory. Hence, 

% we suggest incorporating \textit{prompt optimizers} such as \dspy~\cite{khattab2024dspy}, TextGrad~\cite{yuksekgonul2024textgrad}, and \anthropic~\citep{anthropic_prompt_improver} for refining the effectiveness of the prompts as much as possible.
% for solving the complex problem of repairing FM-synthesized \dsl programs. As such, 
% Hence, we suggest for the application of \textit{advanced prompt optimizers} such as \dspy~\cite{khattab2024dspy}, TextGrad~\cite{yuksekgonul2024textgrad}, and \anthropic~\citep{anthropic_prompt_improver} to refine the utilized prompts even further and to structure them specific to the highly complex nature of our problem. 
\rev{
\subsection*{\practicalimplication{} \textit{4}: Minimal effort is needed to extend the application of \staticanalyzer{} and \correctionmodule to \dsl languages other than \wfdsl.}}

\rev{To further investigate the generalizability of \staticanalyzer{} and \correctionmodule to workflows generated in \dsl{}s other than \wfdsl, we compare our \dsl with other popular workflow generation languages. Specifically, we compare \wfdsl with GitHub Actions and the LangChain Expression Language (LCEL)~\cite{langchainLangChainExpression} to identify their similarities and differences across various aspects, including, but not limited to, syntax and control-flow constructs.}

\rev{As can be observed from \autoref{tab:workflowdsl-vs-others}, \wfdsl  shares close similarities with GitHub Actions and LCEL. With the exception of the primary purpose for generating the workflows and the base language for doing so, across all other aspects, specifically the syntax characteristics and available control flow constructs, \wfdsl shares the same qualities with at least one of the other two \dsl{}s. This means, with the exception of Structure Validator~\bluecircled{A} and Intermediate Representation Constructor~\redcircled{B} in \staticanalyzer{} (Section~\ref{sec:static-analyzer}), that directly operate on \json objects and instantiated \wfdsl programs, all other modules of \staticanalyzer can be generalized to detect defect incidences across GitHub Actions and LCEL workflows.} 

\rev{\begin{table}[!ht]
\centering
\captionsetup{justification=centering}
\caption{Comparison of \wfdsl with GitHub Actions and LCEL}
\label{tab:workflowdsl-vs-others}
\resizebox{\columnwidth}{!}{ %
\begin{tabular}{@{}llll@{}}
\toprule
\multicolumn{1}{c}{\textbf{Aspect}}   & \multicolumn{1}{c}{\textbf{\wfdsl}} & \multicolumn{1}{c}{\textbf{GitHub Actions}} & \multicolumn{1}{c}{\textbf{LCEL}}          \\ \midrule
\textit{Primary Purpose of Workflows} & General-purpose                                    & CI/CD pipeline automation                   & \fm application composition \\
\textit{Base Language}                & JSON                                               & YAML                                        & Python                                     \\
\textit{Syntax Style}                 & Action chaining                                    & Action chaining                             & Action chaining                            \\
\textit{Has Composition?}             & Yes                                                & Yes                                         & Yes                                        \\
\textit{Has Error Handling?}          & Yes                                                & Yes                                         & Yes                                        \\
\textit{Has Conditionals/ Branches?}  & Yes                                                & Yes                                         & Yes                                        \\
\textit{Has Loops?}                   & Yes                                                & Yes                                         & No                                         \\
\textit{Has Parallelism?}             & Yes                                                & Yes                                         & Yes                                        \\
\textit{Artifacts Are Reusable?}      & Yes                                                & Yes                                         & Yes                                        \\
\textit{Type Safety}                  & Schema validation, Pydantic                        & Schema validation                           & Pydantic                                   \\ \bottomrule
\end{tabular}
}%
\end{table}}

\rev{Similarly for \correctionmodule, with the exception of specialized instructions for repairing \badjson{}s and \baddsl{}s, all other prompts and \correctionmodule{}'s feedback mechanism can be applied to repair defective GitHub Actions and LCEL workflows. In other words, we can conclude that other than building specialized intermediate representations for GitHub Actions and LCEL workflows, not much effort is needed to extend the application of \staticanalyzer{} and \correctionmodule to workflows from \dsl{}s other than \wfdsl.}  

\section{Threats to the Validity}
\label{sec:threats}
In this section, we discuss threats to validity across three categories: \textit{construct}, \textit{internal}, and \textit{external} validity.

\subsection{Threats to Construct Validity}
\label{sec: construct-validity}
These threats concern whether our measurements capture the intended concepts.
In the empirical study of defects, the identified \rev{$20$} defect types may still miss rare or context-specific defects.
Although we conduct extensive manual analysis, we acknowledge that the taxonomy may not be fully exhaustive.
Because this is the first systematic effort to categorize these defects, we position the taxonomy as an initial framework that future studies can refine, expand, and validate.

\subsection{Threats to Internal Validity}
\label{sec: Internal-validity}
These threats concern factors that may influence the validity of our procedures and findings.
To build the defect taxonomy, two experts conduct an \opencoding analysis.
As a result, inspector bias may influence defect labeling.
To reduce this risk, the inspectors first label an initial set of \workflow{}s independently, then meet to resolve disagreements and align their interpretations of defect categories.
Prior work reports that this process reduces examiner bias~\citep{masoumzadeh2025experts}.
In cases where the two inspectors do not reach consensus, a third inspector makes the final decision, following recommendations from similar studies~\citep{oliveira2020collaborative, masoumzadeh2025experts}.

To reduce confounding factors in repair evaluation, we assess \correctionmodule for each defect type independently.
However, real-world \workflow{}s can contain multiple interacting defects.
In such cases, repairing one defect can introduce a new defect or hide another one, which can delay or prevent detection and repair.
To mitigate this issue, we apply a test-time iterative strategy that repeatedly executes \correctionmodule until all detectable defect incidences are repaired, including incidences introduced during prior repair attempts.

\subsection{Threats to External Validity}
\label{sec: external-validity}
These threats concern the generalizability of our results.
To derive the taxonomy of defects in FM-\rev{generated} \dsl programs, we conduct \opencoding on a dataset of \rev{$65$} \workflow samples.
This sample may not contain all possible defect types.
To mitigate this risk, we augment observed defect categories with defect types reported in programming languages and software engineering literature.
This allows us to include categories that may not be identifiable through manual inspection alone or without workflow execution, but are observed in related artifacts such as FM-generated general-purpose programs and traditional \dsl programs.

From our taxonomy, only one defect type (i.e., \baddsl) is specific to \workflow{}s because it captures syntax violations tied to a particular \dsl grammar (i.e., \wfdsl).
All remaining defect types can generalize to multi-step processes that are automated using \fms.
In addition, five defect types can apply to single-task processes automated by a single \fm call: over-engineering, incorrect inputs, \badoutput, all sub-categories of incorrect skill usage (i.e., incorrect skills, inefficient skills, \hallucinatedskills, and \badparameter), and incorrect type casting.
\rev{Therefore, as long as a particular DSL grammar can be parsed and intermediate representations (e.g., a CFG) can be constructed, \staticanalyzer and \correctionmodule can be adapted to real-world industrial workflows.}

If the target \dsl grammar changes, some defect types may also change.
However, when building the taxonomy, we consolidate relevant literature to include plausible defect categories, including those that are difficult to detect through manual open coding or prior to workflow execution.
Hence, except for grammar-specific syntax violations, core defect categories are expected to remain largely consistent across grammars.
Their prevalence, however, can vary by \dsl grammar.

\section{Related work}
\label{sec:related_work}

% \keheliya{We need to categorize this into three subsections.} 
% % \cite{ayala2024}
% % \cite{Gandhi2023} 
% \cite{Desai2016}
% % \cite{li2024}
% % \cite{Bassamzadeh2024}
% % \cite{Xu2024}

% \keheliya{
% Some points to discuss
% \begin{itemize}
%     \item human  level interfaces vs low-level interfaces
%     \item idiot proof
%     \item low-resource language problem
%     \item safety rules
%     \item why do you need this? instead of source code
%     \item leveraging MCP - action space
%     \item agent computer interfaces
%     \item safety efficiency
%     \item we're in this part of the world
%     \item smaller in size
%     \item safety will drop
% \end{itemize}
% }

% \keheliya{This should be moved to the end. Just before the conclusion.}

In this section, we situate our work with respect to literature on \dsl program and workflow synthesis by \fms, \fms for program repair, and static analysis of \dsl programs.
\subsection{\fm{}-based Synthesis of \dsl Programs and Workflows}

Desai et al.~\cite{Desai2016} presented a 
% general 
framework for constructing program synthesizers that take natural language inputs and produce expressions in a target \dsl. Their approach constructs these synthesizers using 
three inputs, a \dsl specification, 
% definition,
a training data set, and 
% assistance in the construction of 
a words-to-token dictionary.
% , and then from these inputs constructs a corresponding NL-2-\dsl synthesizer.
Ghandhi et al.~\cite{Gandhi2023} proposed an Office Domain-Specific Language as an \fm{}-friendly \dsl for Office Applications, enabling \fms to accurately translate natural language user queries into \dsl programs interpretable by Office applications. The key idea is that while \fms can produce semantically or syntactically incorrect code, a high-level, \fm{}-friendly language can help mitigate program synthesis errors. However, due to the inherent nature of auto-regressive language models, the synthesized program is not guaranteed to be error-free.

Bassamzadeh and Methani~\cite{Bassamzadeh2024} studied the ability of \fms to transform natural language into \dsl{}s. Their experiments showed that while a fine-tuned \fm scored the best on the code similarity metric, retrieval-augmented generation with optimizations can achieve parity. However, based on the compilation rate, they demonstrated that both settings still got the syntax wrong many times. Similarly, Ayala et al.~\cite{ayala2025fine} compared the performance of fine-tuned small language models (Mistral-Nemo-12B-Base) against propriety (\openai{}'s \gpt family of models and \google{}'s \gemini{}-2.0-Flash) and open-source (\meta{}'s \llama{}-3.3-70B-Instruct) LLMs in synthesizing low-code workflows as structured outputs. Although fine-tuned small language models demonstrated an average 10\% better performance in such synthesis compared to directly prompting LLMs, both settings still generated defective workflows. Barke et al.~\cite{barke2024hysynth} evaluated the performance of \fms in solving structured reasoning and data transformation tasks through Program By Example (PBE), searching for a \dsl that confines the direct transformation from the input to the output of the target reasoning task at hand. They found that even the state-of-the-art proprietary \gpt{}-4o model performs poorly, solving only $10$\% of the tasks in the \textit{Abstraction and Reasoning Corpus}, a benchmark targeted towards human-like structured reasoning tasks. To enable the \fms to synthesize such PBE \dsl{}s successfully, the researchers implemented a correct-by-construction approach that constrains the models' generation by guiding them, using a context-free language model, through the execution of combinatorial search algorithms to find the \dsl that accurately captures the input-output pair transformations.
Liu et al.~\cite{liu2025sew} investigated various workflow representation schemes, ranging from pseudocode to general-purpose scripting languages such as Python and business process languages, including BPMN, to identify the most effective structures for interpretation by LLMs. Despite using different representations and even introducing a complex, multi-agent, self-evolving framework to enhance the structure of synthesized workflows, the researchers found that a number of generated workflows are not executable. Therefore, the implementation of additional workflow-structure validation mechanisms is required.
The results of these studies align with our observation that \fms struggle to synthesize low-resource \dsl programs, underscoring the need to enforce additional guardrails for defect detection and repair.

Several studies~\cite{Zeng2023, Xu2024, ayala2024, xu2024aios} proposed automatic workflow generation using \fms. 
A key observation from all of these approaches is that although some form of functional verification and simulations in real-world workflow execution engines has confirmed the effectiveness of these tools, the specifics regarding syntax errors or semantic defects in their approach remain unclear. In~\cite{zhang2024gh}, researchers demonstrated the effectiveness of \fms in directly synthesizing GitHub Action Workflows. Despite being a form of \dsl, these workflows are readily accessible through GitHub open-source repositories; thus, \fms are not faced with the low-resource-language problem during program synthesis. Furthermore, the scope of these workflows is notably narrower compared to general-purpose workflows, focusing solely on automating the CI/CD stages of the software release pipeline. Therefore, the success of \fms in synthesizing these workflows, in contrast to their limited ability to generate general-purpose workflows, is expected but restricts their use in general-purpose applications.

Based on the limitations of prior research, we recommend a more systematic method (i.e., static analysis), to assess the robustness of such automated workflow generation systems. This will enhance the confidence in employing an \fm{}-based workflow generation system within an enterprise environment.
While prior studies focus on generating \dsl{}s for \fm{}-based applications, which is very similar to our setting, our emphasis in this paper is on detailing fully automatic defect detection and repair after the workflow generation process rather than explaining the \dsl design or workflow generation phase. 

\subsection{Static Analysis of \dsl{}s}

Static analysis for detecting defects is not limited to source code. 
Mandal et al.~\cite{Mandal2018} proposed a generic static analysis framework for \dsl{}s used in safety-critical software, based on abstract interpretation.
Their approach uses data flow analysis and pattern-based matching
techniques to identify domain-specific errors and coding violations for control languages.
Ruiz-Rube et al.~\cite{RuizRube2019} proposed a model-driven interoperability strategy that enables existing source code quality analysis methods and tools to be applied to \dsl artifacts.
To detect anti-patterns in data science pipelines, Rajbahadur et al.~\cite{Rajbahadur2019} analyzed the directed acyclic graph of the pipeline representing its control flow. 
% DAG (Directed Acyclic Graph) representing the control-flow of the pipeline. 

\subsection{\fms for Program Repair}
\fms have 
% recently 
been extensively used in recent studies to repair defects in general-purpose programming languages.
In particular, the field of Automated Program Repair (APR) aims to fix bugs by generating patches.
Hossein et al.~\cite{Hossain2024} proposed \toggle, a program repair framework that combines a bug localization model, an adjustment model to resolve tokenizer inconsistencies, and a bug-fixing model.
Li et al.~\cite{Li2024a} showed that parameter-efficient fine-tuning can be used to reduce computing resource consumption without compromising performance for APR tasks. 
% these APR tasks.
Wadhwa et al.~\cite{wadhwa2024core} introduced \core, an \fm-based APR tool, for generating fixing patches and ranking them, based on human-developer behavior rubric criteria, for assisting developers in code revision and improving code quality issues. \core involves a duo of LLMs, the first one in charge of generating code patches, based on a list of developer-provided recommendations for addressing the issues flagged by a static analyzer, and the latter assigning ranks to generated patches and ordering them based on the ranks.
Batole et al.~\cite{batole2025llm} proposed \localizeagent, a multi-agent framework, specifically designed for executing program analysis tools for localizing design issues and then ranking them based on their relevance in large codebases. Specifically, \localizeagent is designed around two main limitations of \fms: one is their limited context window, which makes analyzing large codebases infeasible, and the other is their inability to understand non-\fm{}-friendly content, such as non-textual modality outputs from program analysis tools.

The code in the existing APR benchmarks may have been included in the training data of existing LLMs, 
making these language models suffer from the threat of ``data leakage'', which in turn leads them to demonstrate misleading optimistic performance metrics on the APR benchmark tasks.
% leading to misleadingly optimistic performance metrics.
Wu et al.~\cite{Wu2024b} introduced \textit{ConDefects}, a complimentary dataset developed 
% as a complement to 
on top of existing ones
% datasets and 
to eliminate the possibility of overlap between existing APR datasets and the training data of LLMs.
% and data leakage.
% such overlap. 
This dataset contains $1,254$ Java faulty programs and $1,625$ Python faulty programs, all of which are sourced from the online competition platform AtCoder.

Chen et al.~\cite{Chen2024} investigated the performance of popular LLMs in handling repository-level repair tasks. They proposed \textit{RepoBugs}, a benchmark comprising $124$ repository-level bugs from open-source repositories. They then used \gpt{}-3.5 to attempt to tackle their newly introduced benchmark tasks. Using a function locating the errors, the repair rate of the used LLM on \textit{RepoBugs} is measured to be only $22.58$\%, considerably diverging from the performance of the same model on function-level bugs as reported in related studies.
% They demonstrate that experiments with \gpt{}-3.5, based on the function where the error is located, reveal that the repair rate on \textit{RepoBugs} is only 22.58\%, diverging from the performance of GPT 3.5 on function-level bugs as reported in related studies.

To standardize the evaluation of APR approaches, Ouyang et al.~\cite{Ouyang2024} conducted a multidimensional evaluation of nine learning-based and three traditional state-of-the-art APR techniques under the same environment and settings. They employed \textit{Defects4J} benchmark and a newly constructed benchmark named \textit{MuBench}, containing $1,700$ artificial bugs generated by mutating bugs in \textit{Defects4J}.
Furthermore, they highlighted the importance of using multi-dimensional metrics, including comparability, plausibility, genuineness, syntactic equivalence, and trivial compiler equivalence metrics. They observed that the LLM-based APR, \alpharepair~\cite{Xia2022}, can correctly fix the largest number of bugs and demonstrates better adaptability and generalizability concerning overfitting issues. While using \fms to repair general-purpose programming languages is common, the effectiveness of applying \fms to repair \dsl programs has not been explored, which we addressed in this work.
\section{Conclusion and Future Work}
\label{sec:conclusion}
In this work, our goal is to improve the reliability and robustness of \workflow{}s. Prior studies have demonstrated that relying solely on the generative capabilities of \fms, even proprietary ones with a considerable number of learnable parameters and complex architectures, is insufficient to directly synthesize such workflows without defects. This, in turn, underscores the need to implement additional guardrails to detect and repair defects in generated workflows.

Although prior research examined the bugs introduced by \fms during code generation and code translation, the defects generated by \fms during NL-2-\dsl translation, such as synthesizing workflows by \fms, have yet to be investigated.

Hence, we start by conducting a manual analysis of \workflow{}s, open-coding them for the types and prevalence of their defect incidences. 
Based on our findings, we introduce \rev{an initial} %the first-ever 
taxonomy of defect 
% categories 
types within these \dsl workflows, applicable not only to \workflow{}s, but to any processes automatable using \fms, which may involve multiple dependent steps or even be comprised of a single task. Our open-coding analysis identifies a high prevalence of defect incidences across \workflow{}s, with \rev{$89.23$\%} of them containing at least one defect incident, which aligns with the prior studies demonstrating the inability of even the most powerful \fms to directly synthesize non-defective workflows.   
% ---that can be automated using \fms.}
% Based on the magnitude of the problem and the necessity that comes with it for solving the issue, 
Given the magnitude of the problem and the necessity of addressing it, as the next step, we propose the application of static analysis in \rev{\fm{}-generated} \dsl program instances
% {}s 
for detecting defects which may have found their way into the synthesized programs, in spite of controlling the \fms to achieve the maximum synthesis accuracy.
% defect detection.
% Moreover, 
Finally, we propose incorporating the results report of the static analyzer as a direct feedback mechanism in a \fm{}-based correction module, to iteratively repair detected defect incidences.
% of static analyzers to guide \fms for an accurate and comprehensive repair of defective \workflow{}s. 
Our study verifies the limited capabilities of \fms in directly \rev{generating} %synthesizing 
\dsl program instances without any defects even when the programs are synthesized iteratively through implementing the test-time \rev{iterative} strategy.
% for iteratively synthesizing these programs is implemented---}% in a single pass, 
The results of our study embody 
% represents 
the initial steps toward validating and repairing these programs with the assistance of established program analysis techniques. 
\bighead{Future Work}
We plan to further improve the detection performance of \staticanalyzer
% . For this purpose, we will 
by incorporating 
% techniques such as 
dynamic analyses and \smt{} solvers, such as \zthree~\citep{de2008z3}, to precisely pinpoint defect incidences arising due to
% as consequences of 
erroneous\rev{,} complex control flow structures and infeasible program paths in \workflow{}s, which are only detectable at runtime.
Despite \correctionmodule{}'s lower than expected accuracy in repairing incidences of some particular defect types, its varying performance across other defect categories, which even maximizes as 
% reaching 
100\% accuracy for repairing \badoutput and \hallucinatedskills, is a key finding for guiding the future research in how to benefit from \fms{}' capabilities for repairing program defects. Specifically, we plan to systematically explore the benefits of incorporating different prompt optimization techniques to increase the effectiveness of \correctionmodule{}'s prompts, dynamically tailoring them towards each defect category and the \fms{}' respective abilities in repairing each specific defect type.
% We also plan to enhance the repair performance of \correctionmodule by optimizing the prompts using automatic approaches. 
Furthermore, we aim to provide more relevant context and more appropriate examples of each defect category to \correctionmodule by using retrieval-augmented generation techniques~\citep{lewis2020rag}. 
% First, we will collect an extensive set of examples corresponding to \fm{}-generated workflows and their different defect incidences. Afterward, we incorporate varied retrieval techniques to choose the most appropriate instances to the workflow under repair from the collection to include them as few-shot examples while prompting the \fms. 
Finally, similar to general-purpose programming languages, we aim to use static analyzers not only for detecting defects in \rev{\fm{}-generated} \dsl programs, but also for detecting any potential incidences of security vulnerabilities~\citep{lipp2022empirical}, performance bottlenecks, and other anti-patterns~\citep{pereira2022code} in them.
% these \dsl programs. 
Our efforts will contribute to achieving robust FM-powered \dsl programs, enabling reliable workflow synthesis and execution in enterprise settings.

% \input{03-background}
% \input{04-methodology}
% \input{05-evaluation-design}
% \input{06-evaluation-results}
% \input{07-discussion}
% \input{08-Threats}
% \input{09-conclusion}
% \appendix
% \input{10-appendix}

\bibliographystyle{ACM-Reference-Format}
\bibliography{99-references}

@misc{ayala2024,
  doi = {10.48550/ARXIV.2412.00239},
  url = {https://arxiv.org/abs/2412.00239},
  author = {Ayala,  Orlando Marquez and Béchard,  Patrice},
  title = {Generating a Low-code Complete Workflow via Task Decomposition and RAG},
  publisher = {arXiv},
  year = {2024},
  copyright = {arXiv.org perpetual,  non-exclusive license}
}

@inproceedings{Desai2016,
  series = {ICSE ’16},
  title = {Program synthesis using natural language},
  url = {http://dx.doi.org/10.1145/2884781.2884786},
  DOI = {10.1145/2884781.2884786},
  booktitle = {Proceedings of the 38th International Conference on Software Engineering},
  publisher = {ACM},
  author = {Desai,  Aditya and Gulwani,  Sumit and Hingorani,  Vineet and Jain,  Nidhi and Karkare,  Amey and Marron,  Mark and R,  Sailesh and Roy,  Subhajit},
  year = {2016},
  month = may,
  pages = {345--356},
  collection = {ICSE ’16}
}

@misc{Gandhi2023,
  doi = {10.48550/ARXIV.2306.03460},
  url = {https://arxiv.org/abs/2306.03460},
  author = {Gandhi,  Apurva and Nguyen,  Thong Q. and Jiao,  Huitian and Steen,  Robert and Bhatawdekar,  Ameya},
  title = {Natural Language Commanding via Program Synthesis},
  publisher = {arXiv},
  year = {2023},
  copyright = {Creative Commons Attribution Non Commercial No Derivatives 4.0 International}
}

@misc{Bassamzadeh2024,
  doi = {10.48550/ARXIV.2407.02742},
  url = {https://arxiv.org/abs/2407.02742},
  author = {Bassamzadeh,  Nastaran and Methani,  Chhaya},
  title = {A Comparative Study of DSL Code Generation: Fine-Tuning vs. Optimized Retrieval Augmentation},
  publisher = {arXiv},
  year = {2024},
  copyright = {Creative Commons Attribution 4.0 International}
}

@inproceedings{Xu2024,
  series = {ASE ’24},
  title = {LLM4Workflow: An LLM-based Automated Workflow Model Generation Tool},
  url = {http://dx.doi.org/10.1145/3691620.3695360},
  DOI = {10.1145/3691620.3695360},
  booktitle = {Proceedings of the 39th IEEE/ACM International Conference on Automated Software Engineering},
  publisher = {ACM},
  author = {Xu,  Jia and Du,  Weilin and Liu,  Xiao and Li,  Xuejun},
  year = {2024},
  month = oct,
  pages = {2394--2398},
  collection = {ASE ’24}
}

@misc{pydantic,
  author       = {Samuel Colvin},
  title        = {Pydantic: Data validation and settings management using Python type annotations},
  year         = {2024},
  url          = {https://docs.pydantic.dev/latest/},
  note         = {Accessed: March 8, 2025}
}

@misc{chen2021evaluating,
      title={Evaluating Large Language Models Trained on Code}, 
      author={Mark Chen and Jerry Tworek and Henry Jun and Qiming Yuan and Henrique P. de Oliveira Pinto and Jared Kaplan and Harri Edwards and Yuri Burda and Nicholas Joseph and Greg Brockman and Alex Ray and Raul Puri and Gretchen Krueger and Michael Petrov and Heewoo Jun and Prafulla Dhariwal and Matthew Knight and Benjamin Chess and John Schulman and Jakob Pachocki and Shen Li and Pawel Klimov and Miljan Martic and Long Ouyang and Neel Alex and Ryan Teehan and Xu Jiang and Igor Babuschkin and Suchir Balaji and Shantanu Jain and William Saunders and Christopher Hesse and Andrew N. Gomez and Tobias Unterthiner and Filip Pavlov and Jeff Clune and Wojciech Zaremba},
      year={2021},
      url={https://arxiv.org/abs/2107.03374}, 
}

@inproceedings{Pan2024,
  series = {ICSE ’24},
  title = {Lost in Translation: A Study of Bugs Introduced by Large Language Models while Translating Code},
  url = {http://dx.doi.org/10.1145/3597503.3639226},
  DOI = {10.1145/3597503.3639226},
  booktitle = {Proceedings of the IEEE/ACM 46th International Conference on Software Engineering},
  publisher = {ACM},
  author = {Pan,  Rangeet and Ibrahimzada,  Ali Reza and Krishna,  Rahul and Sankar,  Divya and Wassi,  Lambert Pouguem and Merler,  Michele and Sobolev,  Boris and Pavuluri,  Raju and Sinha,  Saurabh and Jabbarvand,  Reyhaneh},
  year = {2024},
  month = apr,
  pages = {1–13},
  collection = {ICSE ’24}
}

@inproceedings{Zeng2023,
  series = {ICAIF ’23},
  title = {FlowMind: Automatic Workflow Generation with LLMs},
  url = {http://dx.doi.org/10.1145/3604237.3626908},
  DOI = {10.1145/3604237.3626908},
  booktitle = {4th ACM International Conference on AI in Finance},
  publisher = {ACM},
  author = {Zeng,  Zhen and Watson,  William and Cho,  Nicole and Rahimi,  Saba and Reynolds,  Shayleen and Balch,  Tucker and Veloso,  Manuela},
  year = {2023},
  month = nov,
  pages = {73--81},
  collection = {ICAIF ’23}
}

@inproceedings{Rajbahadur2019,
  title = {Pitfalls Analyzer: Quality Control for Model-Driven Data Science Pipelines},
  url = {http://dx.doi.org/10.1109/MODELS.2019.00-19},
  DOI = {10.1109/models.2019.00-19},
  booktitle = {2019 ACM/IEEE 22nd International Conference on Model Driven Engineering Languages and Systems (MODELS)},
  publisher = {IEEE},
  author = {Rajbahadur,  Gopi Krishnan and Oliva,  Gustavo Ansaldi and Hassan,  Ahmed E. and Dingel,  Juergen},
  year = {2019},
  month = sep,
  pages = {12--22}
}

@misc{xu2024aios,
  doi = {10.48550/ARXIV.2405.06907},
  url = {https://arxiv.org/abs/2405.06907},
  author = {Xu,  Shuyuan and Li,  Zelong and Mei,  Kai and Zhang,  Yongfeng},
  title = {AIOS Compiler: LLM as Interpreter for Natural Language Programming and Flow Programming of AI Agents},
  publisher = {arXiv},
  year = {2024},
  copyright = {arXiv.org perpetual,  non-exclusive license}
}

@inproceedings{srivastava2010program,
  title={From program verification to program synthesis},
  author={Srivastava, Saurabh and Gulwani, Sumit and Foster, Jeffrey S},
  booktitle={Proceedings of the 37th annual ACM SIGPLAN-SIGACT symposium on Principles of programming languages},
  pages={313--326},
  year={2010}
}

@article{xuan2016nopol,
  title={Nopol: Automatic repair of conditional statement bugs in java programs},
  author={Xuan, Jifeng and Martinez, Matias and Demarco, Favio and Clement, Maxime and Marcote, Sebastian Lamelas and Durieux, Thomas and Le Berre, Daniel and Monperrus, Martin},
  journal={IEEE Transactions on Software Engineering},
  volume={43},
  number={1},
  pages={34--55},
  year={2016},
  publisher={IEEE}
}

@article{sun2022consistency,
author = {Sun, Zhihang and Fan, Hongyu and He, Fei},
title = {Consistency-preserving propagation for SMT solving of concurrent program verification},
year = {2022},
issue_date = {October 2022},
publisher = {Association for Computing Machinery},
address = {New York, NY, USA},
volume = {6},
number = {OOPSLA2},
url = {https://doi.org/10.1145/3563321},
doi = {10.1145/3563321},
journal = {Proc. ACM Program. Lang.},
month = oct,
articleno = {158},
numpages = {28}
}

@article{Hossain2024,
  title = {A Deep Dive into Large Language Models for Automated Bug Localization and Repair},
  volume = {1},
  ISSN = {2994-970X},
  url = {http://dx.doi.org/10.1145/3660773},
  DOI = {10.1145/3660773},
  number = {FSE},
  journal = {Proceedings of the ACM on Software Engineering},
  publisher = {Association for Computing Machinery (ACM)},
  author = {Hossain,  Soneya Binta and Jiang,  Nan and Zhou,  Qiang and Li,  Xiaopeng and Chiang,  Wen-Hao and Lyu,  Yingjun and Nguyen,  Hoan and Tripp,  Omer},
  year = {2024},
  month = jul,
  pages = {1471--1493}
}

@inproceedings{Chen2024,
  series = {ICSE-Companion ’24},
  title = {When Large Language Models Confront Repository-Level Automatic Program Repair: How Well They Done?},
  url = {http://dx.doi.org/10.1145/3639478.3647633},
  DOI = {10.1145/3639478.3647633},
  booktitle = {Proceedings of the 2024 IEEE/ACM 46th International Conference on Software Engineering: Companion Proceedings},
  publisher = {ACM},
  author = {Chen,  Yuxiao and Wu,  Jingzheng and Ling,  Xiang and Li,  Changjiang and Rui,  Zhiqing and Luo,  Tianyue and Wu,  Yanjun},
  year = {2024},
  month = apr,
  pages = {459--471},
  collection = {ICSE-Companion ’24}
}

@inproceedings{Li2024a,
  series = {ASE ’24},
  title = {Exploring Parameter-Efficient Fine-Tuning of Large Language Model on Automated Program Repair},
  url = {http://dx.doi.org/10.1145/3691620.3695066},
  DOI = {10.1145/3691620.3695066},
  booktitle = {Proceedings of the 39th IEEE/ACM International Conference on Automated Software Engineering},
  publisher = {ACM},
  author = {Li,  Guochang and Zhi,  Chen and Chen,  Jialiang and Han,  Junxiao and Deng,  Shuiguang},
  year = {2024},
  month = oct,
  pages = {719--731},
  collection = {ASE ’24}
}

@inproceedings{Wu2024b,
  series = {FSE ’24},
  title = {ConDefects: A Complementary Dataset to Address the Data Leakage Concern for LLM-Based Fault Localization and Program Repair},
  url = {http://dx.doi.org/10.1145/3663529.3663815},
  DOI = {10.1145/3663529.3663815},
  booktitle = {Companion Proceedings of the 32nd ACM International Conference on the Foundations of Software Engineering},
  publisher = {ACM},
  author = {Wu,  Yonghao and Li,  Zheng and Zhang,  Jie M. and Liu,  Yong},
  year = {2024},
  month = jul,
  pages = {642--646},
  collection = {FSE ’24}
}

@inproceedings{Ouyang2024,
  series = {ISSTA ’24},
  title = {Benchmarking Automated Program Repair: An Extensive Study on Both Real-World and Artificial Bugs},
  url = {http://dx.doi.org/10.1145/3650212.3652140},
  DOI = {10.1145/3650212.3652140},
  booktitle = {Proceedings of the 33rd ACM SIGSOFT International Symposium on Software Testing and Analysis},
  publisher = {ACM},
  author = {Ouyang,  Yicheng and Yang,  Jun and Zhang,  Lingming},
  year = {2024},
  month = sep,
  pages = {440--452},
  collection = {ISSTA ’24}
}

@inproceedings{Xia2022,
  series = {ESEC/FSE ’22},
  title = {Less training,  more repairing please: revisiting automated program repair via zero-shot learning},
  url = {http://dx.doi.org/10.1145/3540250.3549101},
  DOI = {10.1145/3540250.3549101},
  booktitle = {Proceedings of the 30th ACM Joint European Software Engineering Conference and Symposium on the Foundations of Software Engineering},
  publisher = {ACM},
  author = {Xia,  Chunqiu Steven and Zhang,  Lingming},
  year = {2022},
  month = nov,
  pages = {959–971},
  collection = {ESEC/FSE ’22}
}

@article{Ji2023,
  title = {Survey of Hallucination in Natural Language Generation},
  volume = {55},
  ISSN = {1557-7341},
  url = {http://dx.doi.org/10.1145/3571730},
  DOI = {10.1145/3571730},
  number = {12},
  journal = {ACM Computing Surveys},
  publisher = {Association for Computing Machinery (ACM)},
  author = {Ji,  Ziwei and Lee,  Nayeon and Frieske,  Rita and Yu,  Tiezheng and Su,  Dan and Xu,  Yan and Ishii,  Etsuko and Bang,  Ye Jin and Madotto,  Andrea and Fung,  Pascale},
  year = {2023},
  month = mar,
  pages = {1--38}
}

@inproceedings{yang2024sweagent,
  title={{SWE}-agent: Agent-Computer Interfaces Enable Automated Software Engineering},
  author={John Yang and Carlos E Jimenez and Alexander Wettig and Kilian Lieret and Shunyu Yao and Karthik R Narasimhan and Ofir Press},
  booktitle={The Thirty-eighth Annual Conference on Neural Information Processing Systems},
  year={2024},
  url={https://arxiv.org/abs/2405.15793}
}

@inproceedings{
chen2024teaching,
title={Teaching Large Language Models to Self-Debug},
author={Xinyun Chen and Maxwell Lin and Nathanael Sch{\"a}rli and Denny Zhou},
booktitle={The Twelfth International Conference on Learning Representations {(ICLR)}},
year={2024},
url={https://openreview.net/forum?id=KuPixIqPiq}
}

@inproceedings{
gou2024critic,
title={{CRITIC}: Large Language Models Can Self-Correct with Tool-Interactive Critiquing},
author={Zhibin Gou and Zhihong Shao and Yeyun Gong and yelong shen and Yujiu Yang and Nan Duan and Weizhu Chen},
booktitle={The Twelfth International Conference on Learning Representations  {(ICLR)}},
year={2024},
url={https://openreview.net/forum?id=Sx038qxjek}
}

@misc{langchainLangChainExpression,
  author = {{L}ang{C}hain},
  title = {{L}ang{C}hain {E}xpression {L}anguage ({L}{C}{E}{L})},
  year = {2025},
  url = {https://python.langchain.com/docs/concepts/lcel/},
  note = {Accessed 18-07-2025}
}

@misc{genaiscript,
  author = {Microsoft},
  title = {{G}enerative {A}{I} {S}cripting},
  year = {2025},
  url = {https://microsoft.github.io/genaiscript/},
  note = {Accessed 18-07-2025}
}

@misc{mcp,
  author = {{MCP}},
  title = {Introduction to Model Context Protocol},
  year = {2024},
  url = {https://modelcontextprotocol.io/introduction},
  note = {Accessed: 2025-03-13}
}

@inproceedings{Mandal2018,
  title = {A Generic Static Analysis Framework for Domain-specific Languages},
  url = {http://dx.doi.org/10.1109/ETFA.2018.8502576},
  DOI = {10.1109/etfa.2018.8502576},
  booktitle = {2018 IEEE 23rd International Conference on Emerging Technologies and Factory Automation (ETFA)},
  publisher = {IEEE},
  author = {Mandal,  Avijit and Mohan,  Devina and Jetley,  Raoul and Nair,  Sreeja and D’Souza,  Meenakshi},
  year = {2018},
  month = sep,
  pages = {27--34}
}

@article{RuizRube2019,
  title = {Applying static code analysis for domain-specific languages},
  volume = {19},
  ISSN = {1619-1374},
  url = {http://dx.doi.org/10.1007/s10270-019-00729-w},
  DOI = {10.1007/s10270-019-00729-w},
  number = {1},
  journal = {Software and Systems Modeling},
  publisher = {Springer Science and Business Media LLC},
  author = {Ruiz-Rube,  Iván and Person,  Tatiana and Dodero,  Juan Manuel and Mota,  José Miguel and Sánchez-Jara,  Javier Merchán},
  year = {2019},
  month = apr,
  pages = {95--110}
}

@article{allen1970control,
  title={Control flow analysis},
  author={Allen, Frances E},
  journal={ACM Sigplan Notices},
  volume={5},
  number={7},
  pages={1--19},
  year={1970},
  publisher={ACM New York, NY, USA}
}

@InProceedings{wei2024magicoder,
  title = 	 {Magicoder: Empowering Code Generation with {OSS}-Instruct},
  author =       {Wei, Yuxiang and Wang, Zhe and Liu, Jiawei and Ding, Yifeng and Zhang, Lingming},
  booktitle = 	 {Proceedings of the 41st International Conference on Machine Learning},
  pages = 	 {52632--52657},
  year = 	 {2024},
  volume = 	 {235},
  series = 	 {Proceedings of Machine Learning Research},
  month = 	 {21--27 Jul},
  publisher =    {PMLR},
  url = 	 {https://proceedings.mlr.press/v235/wei24h.html}
}

@article{luo2023wizardcoder,
  title={WizardCoder: Empowering Code Large Language Models with Evol-Instruct},
  author={Luo, Ziyang and Xu, Can and Zhao, Pu and Sun, Qingfeng and Geng, Xiubo and Hu, Wenxiang and Tao, Chongyang and Ma, Jing and Lin, Qingwei and Jiang, Daxin},
  journal={arXiv preprint arXiv:2306.08568},
  year={2023}
}

@misc{starcoder2,
  doi = {10.48550/ARXIV.2402.19173},
  author = {Lozhkov,  Anton and Li,  Raymond and Allal,  Loubna Ben and Cassano,  Federico and Lamy-Poirier,  Joel and Tazi,  Nouamane and Tang,  Ao and Pykhtar,  Dmytro and Liu,  Jiawei and Wei,  Yuxiang and Liu,  Tianyang and Tian,  Max and Kocetkov,  Denis and Zucker,  Arthur and Belkada,  Younes and Wang,  Zijian and Liu,  Qian and Abulkhanov,  Dmitry and Paul,  Indraneil and Li,  Zhuang and Li,  Wen-Ding and Risdal,  Megan and Li,  Jia and Zhu,  Jian and Zhuo,  Terry Yue and Zheltonozhskii,  Evgenii and Dade,  Nii Osae Osae and Yu,  Wenhao and Krauß,  Lucas and Jain,  Naman and Su,  Yixuan and He,  Xuanli and Dey,  Manan and Abati,  Edoardo and Chai,  Yekun and Muennighoff,  Niklas and Tang,  Xiangru and Oblokulov,  Muhtasham and Akiki,  Christopher and Marone,  Marc and Mou,  Chenghao and Mishra,  Mayank and Gu,  Alex and Hui,  Binyuan and Dao,  Tri and Zebaze,  Armel and Dehaene,  Olivier and Patry,  Nicolas and Xu,  Canwen and McAuley,  Julian and Hu,  Han and Scholak,  Torsten and Paquet,  Sebastien and Robinson,  Jennifer and Anderson,  Carolyn Jane and Chapados,  Nicolas and Patwary,  Mostofa and Tajbakhsh,  Nima and Jernite,  Yacine and Ferrandis,  Carlos Muñoz and Zhang,  Lingming and Hughes,  Sean and Wolf,  Thomas and Guha,  Arjun and von Werra,  Leandro and de Vries,  Harm},
  title = {StarCoder 2 and The Stack v2: The Next Generation},
  publisher = {arXiv},
  year = {2024},
  copyright = {Creative Commons Attribution Share Alike 4.0 International}
}

@inproceedings{ma-etal-2023-chain,
    title = "Chain-of-Skills: A Configurable Model for Open-Domain Question Answering",
    author = "Ma, Kaixin  and
      Cheng, Hao  and
      Zhang, Yu  and
      Liu, Xiaodong  and
      Nyberg, Eric  and
      Gao, Jianfeng",
    booktitle = "Proceedings of the 61st Annual Meeting of the Association for Computational Linguistics (Volume 1: Long Papers)",
    month = jul,
    year = "2023",
    address = "Toronto, Canada",
    publisher = "Association for Computational Linguistics",
    url = "https://aclanthology.org/2023.acl-long.89",
    doi = "10.18653/v1/2023.acl-long.89",
    pages = "1599--1618",
}

@inproceedings{
zhang2024endtoend,
title={End-to-End Beam Retrieval for Multi-Hop Question Answering},
author={Jiahao Zhang and Haiyang Zhang and Dongmei Zhang and Yong Liu and Shen Huang},
booktitle={2024 Annual Conference of the North American Chapter of the Association for Computational Linguistics},
year={2024},
url={https://arxiv.org/abs/2308.08973}
}

@inproceedings{shi2023,
author = {Shi, Freda and Chen, Xinyun and Misra, Kanishka and Scales, Nathan and Dohan, David and Chi, Ed and Sch\"{a}rli, Nathanael and Zhou, Denny},
title = {Large language models can be easily distracted by irrelevant context},
year = {2023},
publisher = {JMLR.org},
booktitle = {Proceedings of the 40th International Conference on Machine Learning},
articleno = {1291},
numpages = {18},
location = {Honolulu, Hawaii, USA},
series = {ICML'23}
}

@inproceedings{nguyen2013semfix,
  author    = {Hoang Duong Thien Nguyen and Dawei Qi and Abhik Roychoudhury and Satish Chandra},
  title     = {SemFix: Program Repair via Semantic Analysis},
  booktitle = {Proceedings of the 35th International Conference on Software Engineering (ICSE)},
  year      = {2013},
  pages     = {772--781},
  publisher = {IEEE},
  doi       = {10.1109/ICSE.2013.6606613}
}

@inproceedings{de2008z3,
  author    = {Leonardo De Moura and Nikolaj Bj{\o}rner},
  title     = {Z3: An Efficient SMT Solver},
  booktitle = {Proceedings of the 14th International Conference on Tools and Algorithms for the Construction and Analysis of Systems (TACAS)},
  year      = {2008},
  pages     = {337--340},
  publisher = {Springer},
  doi       = {10.1007/978-3-540-78800-3_24}
}

@article{brown2020language,
  author    = {Tom B. Brown and Benjamin Mann and Nick Ryder and Melanie Subbiah and Jared D. Kaplan and Prafulla Dhariwal and Arvind Neelakantan and Pranav Shyam and Girish Sastry and Amanda Askell and others},
  title     = {Language Models are Few-Shot Learners},
  journal   = {Advances in Neural Information Processing Systems (NeurIPS)},
  volume    = {33},
  pages     = {1877--1901},
  year      = {2020},
  publisher = {Curran Associates, Inc.},
  doi       = {10.48550/arXiv.2005.14165},
  archivePrefix = {arXiv},
  eprint    = {2005.14165},
  primaryClass = {cs.CL}
}

@misc{austin2021program,
  title={Program synthesis with large language models},
  author={Austin, Jacob and Odena, Augustus and Nye, Maxwell and Bosma, Maarten and Michalewski, Henryk and Dohan, David and Jiang, Ellen and Cai, Carrie and Terry, Michael and Le, Quoc and others},
  howpublished={arXiv preprint arXiv:2108.07732},
  year={2021}
}

@inproceedings{ahmed2022few,
  title={Few-shot training llms for project-specific code-summarization},
  author={Ahmed, Toufique and Devanbu, Premkumar},
  booktitle={Proceedings of the 37th IEEE/ACM international conference on automated software engineering},
  publisher={Association for Computing Machinery},
  address={New York, NY, USA},
  pages={1--5},
  year={2022}
}

@inproceedings{khattab2024dspy,
  title={DSPy: Compiling Declarative Language Model Calls into Self-Improving Pipelines},
  author={Khattab, Omar and Singhvi, Arnav and Maheshwari, Paridhi and Zhang, Zhiyuan and Santhanam, Keshav and Vardhamanan, Sri and Haq, Saiful and Sharma, Ashutosh and Joshi, Thomas T. and Moazam, Hanna and Miller, Heather and Zaharia, Matei and Potts, Christopher},
  journal={The Twelfth International Conference on Learning Representations},
  year={2024}
}

@misc{anthropic_prompt_improver,
  title        = {Use our prompt improver to optimize your prompts},
  author       = {Anthropic},
  year         = {2025},
  url          = {https://docs.anthropic.com/en/docs/build-with-claude/prompt-engineering/prompt-improver},
  note         = {Accessed: 2025-03-14}
}

@article{yuksekgonul2024textgrad,
      title={TextGrad: Automatic "Differentiation" via Text},
      author={Mert Yuksekgonul and Federico Bianchi and Joseph Boen and Sheng Liu and Zhi Huang and Carlos Guestrin and James Zou},
      year={2024},
      eprint={2406.07496},
      archivePrefix={arXiv}
}

@inproceedings{wen2024enchanting,
  title={Enchanting program specification synthesis by large language models using static analysis and program verification},
  author={Wen, Cheng and Cao, Jialun and Su, Jie and Xu, Zhiwu and Qin, Shengchao and He, Mengda and Li, Haokun and Cheung, Shing-Chi and Tian, Cong},
  booktitle={International Conference on Computer Aided Verification},
  pages={302--328},
  year={2024},
  organization={Springer}
}

@article{oliveira2020collaborative,
  title={Collaborative or individual identification of code smells? On the effectiveness of novice and professional developers},
  author={Oliveira, Roberto and de Mello, Rafael and Fernandes, Eduardo and Garcia, Alessandro and Lucena, Carlos},
  journal={Information and Software Technology},
  volume={120},
  year={2020},
  publisher={Elsevier}
}

@inproceedings{lewis2020rag,
author = {Lewis, Patrick and Perez, Ethan and Piktus, Aleksandra and Petroni, Fabio and Karpukhin, Vladimir and Goyal, Naman and K\"{u}ttler, Heinrich and Lewis, Mike and Yih, Wen-tau and Rockt\"{a}schel, Tim and Riedel, Sebastian and Kiela, Douwe},
title = {Retrieval-augmented generation for knowledge-intensive NLP tasks},
year = {2020},
isbn = {9781713829546},
booktitle = {Proceedings of the 34th International Conference on Neural Information Processing Systems},
}

@article{pereira2022code,
  title={Code smells detection and visualization: a systematic literature review},
  author={Pereira dos Reis, Jos{\'e} and Brito e Abreu, Fernando and de Figueiredo Carneiro, Glauco and Anslow, Craig},
  journal={Archives of Computational Methods in Engineering},
  volume={29},
  number={1},
  pages={47--94},
  year={2022},
  publisher={Springer}
}

@inproceedings{lipp2022empirical,
  title={An empirical study on the effectiveness of static C code analyzers for vulnerability detection},
  author={Lipp, Stephan and Banescu, Sebastian and Pretschner, Alexander},
  booktitle={Proceedings of the 31st ACM SIGSOFT international symposium on software testing and analysis},
  pages={544--555},
  year={2022}
}

@misc{benaich2024state,
  title={State of AI report},
  author={Benaich, Nathan and Hogarth, Ian},
  url = {https://www.stateof.ai/},
  year={2024}
}

@book{russell2016workflow,
  title={Workflow patterns: the definitive guide},
  author={Russell, Nick and Van Der Aalst, Wil Mp and Ter Hofstede, Arthur HM},
  year={2016},
  publisher={Mit Press}
}

@misc{hassan2024rethinking,
  title={{Rethinking software engineering in the foundation model era: From task-driven AI copilots to goal-driven AI pair programmers}},
  author={Hassan, Ahmed E and Oliva, Gustavo A and Lin, Dayi and Chen, Boyuan and Ming, Zhen and others},
  howpublished={arXiv preprint arXiv:2404.10225},
  year={2024}
}

@misc{joel2024survey,
  title={A survey on llm-based code generation for low-resource and domain-specific programming languages},
  author={Joel, Sathvik and Wu, Jie JW and Fard, Fatemeh H},
  howpublished={arXiv preprint arXiv:2410.03981},
  year={2024}
}

@article{Wu2025,
  title = {HumanEvalComm: Benchmarking the Communication Competence of Code Generation for LLMs and LLM Agent},
  ISSN = {1557-7392},
  url = {http://dx.doi.org/10.1145/3715109},
  DOI = {10.1145/3715109},
  journal = {ACM Transactions on Software Engineering and Methodology},
  publisher = {Association for Computing Machinery (ACM)},
  author = {Wu,  Jie Jw and Fard,  Fatemeh H.},
  year = {2025},
  month = jan 
}

@inproceedings{zhang2024gh,
author = {Zhang, Xinyu and Muralee, Siddharth and Cherupattamoolayil, Sourag and Machiry, Aravind},
title = {On the Effectiveness of Large Language Models for GitHub Workflows},
year = {2024},
publisher = {Association for Computing Machinery},
address = {New York, NY, USA},
doi = {10.1145/3664476.3664497},
booktitle = {Proceedings of the 19th International Conference on Availability, Reliability and Security},
articleno = {32},
numpages = {14},
series = {ARES '24}
}

@inproceedings{wang2024sotopia,
    title = "{SOTOPIA}-{\ensuremath{\pi}}: Interactive Learning of Socially Intelligent Language Agents",
    author = "Wang, Ruiyi  and
      Yu, Haofei  and
      Zhang, Wenxin  and
      Qi, Zhengyang  and
      Sap, Maarten  and
      Bisk, Yonatan  and
      Neubig, Graham  and
      Zhu, Hao",
    booktitle = "Proceedings of the 62nd Annual Meeting of the Association for Computational Linguistics (Volume 1: Long Papers)",
    year = "2024",
    publisher = "Association for Computational Linguistics",
    doi = "10.18653/v1/2024.acl-long.698",
    pages = "12912--12940",
}

@inproceedings{abbasiantaeb2024let,
  title={Let the llms talk: Simulating human-to-human conversational qa via zero-shot llm-to-llm interactions},
  author={Abbasiantaeb, Zahra and Yuan, Yifei and Kanoulas, Evangelos and Aliannejadi, Mohammad},
  booktitle={Proceedings of the 17th ACM International Conference on Web Search and Data Mining},
  publisher={Association for Computing Machinery},
  address={New York, NY, USA},
  pages={8--17},
  year={2024}
}

@article{kamath2024scope,
  title={Scope ambiguities in large language models},
  author={Kamath, Gaurav and Schuster, Sebastian and Vajjala, Sowmya and Reddy, Siva},
  journal={Transactions of the Association for Computational Linguistics},
  volume={12},
  pages={738--754},
  year={2024},
}

@misc{openai2024prompt,
  author       = {{OpenAI}},
  title        = {Text Generation and Prompting},
  year         = {2024},
  url          = {https://platform.openai.com/docs/guides/text?api-mode=responses},
  note         = {Accessed: 2025-06-29}
}

@misc{anthropic2024prompt,
  author       = {{Anthropic}},
  title        = {Prompt Engineering Overview},
  year         = {2024},
  url          = {https://docs.anthropic.com/en/docs/build-with-claude/prompt-engineering/overview},
  note         = {Accessed: 2025-06-29}
}

@misc{xia2024agentless,
  title={Agentless: Demystifying llm-based software engineering agents},
  author={Xia, Chunqiu Steven and Deng, Yinlin and Dunn, Soren and Zhang, Lingming},
  howpublished={arXiv preprint arXiv:2407.01489},
  year={2024}
}

@inproceedings{
wang2025openhands,
title={OpenHands: An Open Platform for {AI} Software Developers as Generalist Agents},
author={Xingyao Wang and Boxuan Li and Yufan Song and Frank F. Xu and Xiangru Tang and Mingchen Zhuge and Jiayi Pan and Yueqi Song and Bowen Li and Jaskirat Singh and Hoang H. Tran and Fuqiang Li and Ren Ma and Mingzhang Zheng and Bill Qian and Yanjun Shao and Niklas Muennighoff and Yizhe Zhang and Binyuan Hui and Junyang Lin and Robert Brennan and Hao Peng and Heng Ji and Graham Neubig},
booktitle={The Thirteenth International Conference on Learning Representations},
year={2025},
url={https://openreview.net/forum?id=OJd3ayDDoF}
}

@inproceedings{jasinski2022natural,
  title={Natural language processing applied to dynamic workflow generation for network management},
  author={Jasinski, Andrzej and Qiao, Yuansong and Fallon, Enda and Flynn, Ronan},
  booktitle={NOMS 2022-2022 IEEE/IFIP Network Operations and Management Symposium},
  pages={1--6},
  year={2022},
  organization={IEEE}
}

@inproceedings{vinci2024repairing,
  title={Repairing Process Models Through Simulation and Explainable AI},
  author={Vinci, Francesco and de Leoni, Massimiliano},
  booktitle={International Conference on Business Process Management},
  pages={129--145},
  year={2024},
  organization={Springer}
}

@misc{wang2024large,
  title={Where Do Large Language Models Fail When Generating Code?},
  author={Wang, Zhijie and Zhou, Zijie and Song, Da and Huang, Yuheng and Chen, Shengmai and Ma, Lei and Zhang, Tianyi},
  howpublished={arXiv preprint arXiv:2406.08731},
  year={2024}
}

@article{wadhwa2024core,
  title={Core: Resolving code quality issues using llms},
  author={Wadhwa, Nalin and Pradhan, Jui and Sonwane, Atharv and Sahu, Surya Prakash and Natarajan, Nagarajan and Kanade, Aditya and Parthasarathy, Suresh and Rajamani, Sriram},
  journal={Proceedings of the ACM on Software Engineering},
  volume={1},
  number={FSE},
  pages={789--811},
  year={2024},
  publisher={ACM New York, NY, USA}
}

@inproceedings{batole2025llm,
  title={An LLM-Based Agent-Oriented Approach for Automated Code Design Issue Localization},
  author={Batole, Fraol and OBrien, David and Nguyen, Tien N and Dyer, Robert and Rajan, Hridesh},
  booktitle={2025 IEEE/ACM 47th International Conference on Software Engineering (ICSE)},
  publisher={IEEE Press},
  pages={637--637},
  year={2025},
  organization={IEEE Computer Society}
}

@misc{ayala2025fine,
  title={Fine-Tune an SLM or Prompt an LLM? The Case of Generating Low-Code Workflows},
  author={Ayala, Orlando Marquez and Bechard, Patrice and Chen, Emily and Baird, Maggie and Chen, Jingfei},
  howpublished={arXiv preprint arXiv:2505.24189},
  year={2025}
}

@article{masoumzadeh2025experts,
  title={Do Experts Agree About Smelly Infrastructure?},
  author={Masoumzadeh, Sogol and Saavedra, Nuno and Maipradit, Rungroj and Wei, Lili and Ferreira, Jo{\~a}o F and Varr{\'o}, D{\'a}niel and McIntosh, Shane},
  journal={IEEE Transactions on Software Engineering},
  year={2025},
  publisher={IEEE}
}

@misc{huang2023large,
  title={Large language models cannot self-correct reasoning yet},
  author={Huang, Jie and Chen, Xinyun and Mishra, Swaroop and Zheng, Huaixiu Steven and Yu, Adams Wei and Song, Xinying and Zhou, Denny},
  howpublished={arXiv preprint arXiv:2310.01798},
  year={2023}
}

@article{kim2023language,
  title={Language models can solve computer tasks},
  author={Kim, Geunwoo and Baldi, Pierre and McAleer, Stephen},
  journal={Advances in Neural Information Processing Systems},
  volume={36},
  pages={39648--39677},
  year={2023}
}

@article{barke2024hysynth,
  title={Hysynth: Context-free llm approximation for guiding program synthesis},
  author={Barke, Shraddha and Anaya Gonzalez, Emmanuel and Kasibatla, Saketh Ram and Berg-Kirkpatrick, Taylor and Polikarpova, Nadia},
  journal={Advances in Neural Information Processing Systems},
  volume={37},
  pages={15612--15645},
  year={2024}
}

@article{mu2024clarifygpt,
  title={Clarifygpt: A framework for enhancing llm-based code generation via requirements clarification},
  author={Mu, Fangwen and Shi, Lin and Wang, Song and Yu, Zhuohao and Zhang, Binquan and Wang, ChenXue and Liu, Shichao and Wang, Qing},
  journal={Proceedings of the ACM on Software Engineering},
  volume={1},
  number={FSE},
  pages={2332--2354},
  year={2024},
  publisher={ACM New York, NY, USA}
}

@misc{gallaba2025towards,
  title={Towards Conversational Development Environments: Using Theory-of-Mind and Multi-Agent Architectures for Requirements Refinement},
  author={Gallaba, Keheliya and Arabat, Ali and Lin, Dayi and Sayagh, Mohammed and Hassan, Ahmed E},
  howpublished={arXiv preprint arXiv:2505.20973},
  year={2025}
}

@article{liu2025sew,
  title={SEW: Self-Evolving Agentic Workflows for Automated Code Generation},
  author={Liu, Siwei and Fang, Jinyuan and Zhou, Han and Wang, Yingxu and Meng, Zaiqiao},
  journal={arXiv preprint arXiv:2505.18646},
  year={2025}
}

@misc{shorten2024structuredrag,
  title={{StructuredRAG: JSON Response Formatting with Large Language Models}},
  author={Shorten, Connor and Pierse, Charles and Smith, Thomas Benjamin and Cardenas, Erika and Sharma, Akanksha and Trengrove, John and van Luijt, Bob},
  howpublished={arXiv preprint arXiv:2408.11061},
  year={2024}
}

@inproceedings{CodeRL2022Le,
author = {Le, Hung and Wang, Yue and Gotmare, Akhilesh Deepak and Savarese, Silvio and Hoi, Steven C.H.},
title = {CodeRL: Mastering Code Generation Through Pretrained Models and Deep Reinforcement Learning},
year = {2022},
isbn = {9781713871088},
booktitle = {Proceedings of the 36th International Conference on Neural Information Processing Systems},
articleno = {1549},
numpages = {15},
series = {NIPS '22}
}

@article{lyu2025top,
  title={Top Pass: Improve Code Generation by Pass@ k-maximized Code Ranking},
  author={Lyu, Zhicun and Li, Xinye and Xie, Zheng and Li, Ming},
  journal={Frontiers of Computer Science},
  volume={19},
  number={8},
  pages={198341},
  year={2025},
}

@inproceedings{openagi,
 author = {Ge, Yingqiang and Hua, Wenyue and Mei, Kai and ji, jianchao and Tan, Juntao and Xu, Shuyuan and Li, Zelong and Zhang, Yongfeng},
 booktitle = {Advances in Neural Information Processing Systems},
 pages = {5539--5568},
 title = {OpenAGI: When LLM Meets Domain Experts},
 volume = {36},
 year = {2023}
}

@misc{openai_pricing,
  author = {{OpenAI}},
  title = {Pricing},
  howpublished = {\url{https://platform.openai.com/docs/pricing}},
  note = {Accessed: 2026-01-10},
  year = {2025}
}

@inproceedings{li-etal-2024-evaluating-mathematical,
    title = "Evaluating Mathematical Reasoning of Large Language Models: A Focus on Error Identification and Correction",
    author = "Li, Xiaoyuan  and
      Wang, Wenjie  and
      Li, Moxin  and
      Guo, Junrong  and
      Zhang, Yang  and
      Feng, Fuli",
    booktitle = "Findings of the Association for Computational Linguistics: ACL 2024",
    year = "2024",
    pages = "11316--11360",
}

@article{cohen1960kappa,
  title={A coefficient of agreement for nominal scales},
  author={Cohen, Jacob},
  journal={Educational \& psychological measurement},
  volume={20},
  number={1},
  year={1960}, 
}

@article{mchugh2012interrater,
  title={Interrater reliability: the kappa statistic},
  author={McHugh, Mary L},
  journal={Biochemia Medica},
  year={2012},
  volume={22},
  pages={276--282},
  url={https://api.semanticscholar.org/CorpusID:5421278}
}

\end{document}